\documentclass[twocolumn]{aastex62}
\usepackage{amsmath,amstext}
\usepackage[T1]{fontenc}
\usepackage{natbib}
\usepackage{sidecap}
\usepackage{longtable}

\begin{document}

\title{Differences between Stable and Unstable Architectures of Compact Planetary Systems} 
\shorttitle{Stable versus Unstable Compact Planetary Architectures}

\author[0000-0001-8736-236X]{Kathryn Volk}
\correspondingauthor{Kathryn Volk}
\email{kat.volk@gmail.com}
\affiliation{Lunar and Planetary Laboratory, The University of Arizona, 1629 E University Blvd, Tucson, AZ 85721}
\affiliation{Planetary Science Institute, 1700 East Fort Lowell, Suite 106, Tucson, AZ 85719, USA}

\author[0000-0002-1226-3305]{Renu Malhotra}
\affil{Lunar and Planetary Laboratory, The University of Arizona, 1629 E University Blvd, Tucson, AZ 85721}

\begin{abstract}
We present a stability analysis of a large set of simulated planetary systems of three or more planets based on architectures of multiplanet systems discovered by \textit{Kepler} and \textit{K2}. 
We propagated 21,400 simulated planetary systems up to 5 billion orbits of the innermost planet; approximately 13\% of these simulations ended in a planet-planet collision within that timespan.
We examined trends in dynamical stability based on dynamical spacings, orbital period ratios, and mass ratios of nearest-neighbor planets as well as the system-wide planet mass distribution and the spectral fraction describing the system's short-term evolution.
We find that instability is more likely in planetary systems with adjacent planet pairs that have period ratios less than two and in systems of greater variance of planet masses. 
Systems with planet pairs at very small dynamical spacings (less than $\sim10-12$ mutual Hill radius) are also prone to instabilities, but instabilities also occur at much larger planetary separations.
We find that a large spectral fraction (calculated from short integrations) is a reasonable predictor of longer-term dynamical instability; systems that have a large number of Fourier components in their eccentricity vectors are prone to secular chaos and subsequent eccentricity growth and instabilities.
\end{abstract}

\keywords{Exoplanets, Exoplanet dynamics, Exoplanet systems, Orbital evolution}

\section{Introduction} \label{sec:intro}

Exoplanet surveys have revealed a large number and wide variety of  multi-planet systems in the solar neighborhood. 
The close orbital spacings of many of the observed multi-planet systems seemingly place many of them on the edge of dynamical stability, despite the gigayear ages of many of their host stars (e.g., as previously discussed in the context of the surprisingly packed {\it Kepler}-11 system; \citealt{Lissauer:2011}).
The study of planetary system stability is a long-standing area of celestial mechanics research (see, e.g., recent discussions in \citealt{Petit:2020,Tamayo:2021,Weiss:2023}). 
The stark contrast between the observed {\it Kepler} multiplanet systems and our own solar system has also sparked many investigations into different models of planetary system formation and migration (see recent reviews by \citealt{Weiss:2023,Lissauer:2023}). 
The statistical distribution of planet masses and orbits based on dynamical stability have also been investigated \citep[e.g.,][]{Malhotra:2015,Tremaine:2015}.
In this paper, we use the observational data of multi-planet system architectures as the basis for a statistical study of planetary system architectures with a focus on two questions: What are the distinguishable features of a stable planetary system architecture? What are the drivers of dynamical instabilities in systems whose architectures are broadly similar to those observed?

To answer these questions, we first identify a few parameters of planetary systems that broadly quantify their architectures, and then we examine how long term stability depends upon small variations of these parameters (within the uncertainties of the observational data of the {\it Kepler} sample). 
The parameters we identify are the following: the dynamical spacings, orbital period ratios, and mass ratios of nearest-neighbor planets; the coefficient of variation in system-wide planet masses; and the spectral fraction describing the system's secular orbital evolution. 
These parameters have been investigated in a few recent studies, though mostly in isolation. 
For example, planetary dynamical spacings have been explored in randomly-generated simulated systems by \cite{Pu:2015,Obertas:2017,Rice:2023}, period ratios have been discussed by \cite{Steffen:2015}, planet mass distributions have been discussed by \cite{Mishra:2023}, and the spectral index of secular orbital evolution has been discussed by  \cite{Volk:2020}. 
In the present work, we examine all of these parameters for a large set of planetary architectures modeled on observed systems.
In addition to expanding our understanding of the features of long-term stable planetary system architectures in general, our study can help identify observed exoplanet systems where dynamical considerations can provide useful constraints on planetary properties or even identify cases where some observational constraints may be spurious. 

In Section~\ref{sec:simulations}, we describe our starting set of observed planetary architectures and the large suite of numerical simulations we performed to investigate system stability.
Section~\ref{sec:analysis} describes our statistical exploration of stability in these simulations as a function of various system parameters.
Section~\ref{sec:individual-systems} provides a brief discussion of a number of notable observed multi-planet systems that warrant additional future dynamical investigations to improve constraints on planetary masses.
Section~\ref{sec:discussion} provides a discussion of our results, focusing on the system-wide properties that are most predictive of stability/instability.
We summarize our findings in Section~\ref{sec:conclusions}.

\section{A large suite of planetary system simulations} \label{sec:simulations}

We initiated this study in 2019 with the then-known set of {\it Kepler} and {\it K2} planetary systems hosting three or more confirmed planets and having a published estimated stellar mass in the NASA Exoplanet Archive\footnote{\url{https://exoplanetarchive.ipac.caltech.edu}} composite data table \citep{composite-table}. 
Subsequently we added all the additional reported systems with four or more planets as of January 2023 \citep{PSCompPars}. 
We also added new simulations of any of our original systems with additional planets reported between 2019 and January 2023.
The resulting 165 multi-planet {\it Kepler} and {\it K2} systems are listed in Table~\ref{tab:systems} in Appendix~\ref{a:systems}, including  citations to published planet radius, planet mass, planet eccentricity, and stellar mass estimates we make use of. 
We do not include circumbinary planets in our sample, but we do include 8 planetary systems that orbit one star within likely or confirmed multi-star systems. 
These systems are noted in Table~\ref{tab:systems}.  
For these systems, we do not account for the stellar companions in our modeling as we are not attempting high precision orbital propagation, but investigating only broad dynamical stability properties of their orbital architectures.

Our sample of planetary systems includes 165 systems with three or more transiting planets; a majority of these planets have radii smaller than $4R_\earth$ and only a minority have measured masses. 
For 162 of these systems, we run a set of simulations with planetary masses assigned from a prescribed (statistical) mass-radius relationship. 
Henceforth, we will refer to this simulation set as the ``MR relationship cases''.
Our sample includes 47 systems with published mass and/or eccentricity constraints from either radial velocity (RV) or transit-timing variations (TTV) for at least one planet that provide mass estimates more stringent than a statistical MR relationship. 
For these, we run a set of simulations using those mass/eccentricity constraints. 
Henceforth, we will refer to this simulation set as the ``TTV/RV cases''; note that 42 of the 45 systems in this set are also in the MR-relationship set.
Full details of the initialization of each of these cases are given below.
We emphasize that none of our simulations represent the \textit{exact} configuration of the observed planetary systems.
Our goal is to probe general trends in planetary system stability based on system-wide properties such as period ratio distributions and typical planetary mass distributions.

\subsection{MR relationship cases}\label{sec:MR-ic}

For the 162 systems with transiting planets smaller than $4R_{\earth}$, we initialize 100 simulation instances as follows using the same procedures as in \cite{Volk:2020} (which presented results from a subset of the wider range of simulations in this work). 
We assign individual planetary masses by randomly sampling each planet's radius uniformly within the reported 1$\sigma$ uncertainties and then use that radius to sample a mass from the statistical MR relationship from \cite{Wolfgang:2016}\footnote{Their code is available at \url{github.com/dawolfgang/MRrelation}}. 
That mass-radius relationship can be sampled based on either RV or TTV priors, so we evenly split our simulations between those two cases.
For the small number of systems that have some planets whose radii exceed $4R_{\earth}$ (the size range over which \cite{Wolfgang:2016}'s MR relationship is valid), we revert to TTV or RV mass constraints if they exist or to the Exoplanet Archive's listed mass estimate based on the MR relationship from \cite{Chen:2017}\footnote{See \url{ https://exoplanetarchive.ipac.caltech.edu/docs/composite_calc.html}}. 
These special cases are noted in Table~\ref{tab:systems}.
We then assign stellar masses and planet orbital periods chosen randomly within the $1\sigma$ error estimates.
We choose to sample all the parameter uncertainties uniformly and only within 1-$\sigma$ because available computational and time resources limit us to a relatively small number of simulations.
This allows us to probe a fair range of parameters with reasonable density of initial conditions; we are also not attempting to reproduce the exact observed systems.
We assign small initial orbital eccentricities $e$ and inclinations $i$ from Rayleigh distributions of widths $\sigma_e=0.02$ and $\sigma_i=1.4^\circ$, which are consistent with estimates of the intrinsic eccentricity and inclination distributions for {\it Kepler} multiplanet systems \citep[e.g.,][]{Hadden:2014, Xie:2016,VanEylen:2015}. 
We randomize the planets' longitudes of ascending node $\Omega$, arguments of perihelion $\omega$, and mean anomalies $M$. 
This resulted in 16,500 simulated systems representing 162 unique observed multiplanet systems. 
Three observed systems were included twice: K2-138 due to including additional planets detected since our initial exploration of that system architecture,  Kepler-90 for an additional choice of planetary mass assumptions for the $R>4R_\earth$ planets, and Kepler-1542 for significantly updated planetary radius measurements. 
Each instance was integrated forward using the {\sc{mercurius}} integrator within {\sc{rebound}} \citep{Rein:2012}, which uses the  {\sc{whfast}} \citep{whfast} integrator for most timesteps (using a timestep equal to 1/40th of the shortest orbital period) and switches to the adaptive step-size {\sc{ias15}} \citep{ias15} integrator to resolve close-encounters between massive bodies.
We include a first-order post-Newtonian general relativity correction as a user-defined force based on a simplified potential \citep[the implementation is described in][]{2020MNRAS.491.2885T}.
We integrated each simulated system until a planetary collision was detected or the integration reached $5\times10^9$ orbital periods of the innermost planet. As discussed in \cite{Volk:2015}, ejection is highly unlikely for planets orbiting as close to their stars as the observed transiting multiplanet systems and stellar collisions are also expected to be rare; while our simulations did have inner and outer boundary conditions, these were never triggered, and all simulations either survived the maximum integration length or ended in planetary collisions.

\subsection{TTV/RV cases}\label{sec:non-MR-ic}

For the 47 observed systems with at least one planet with a reasonably well-constrained TTV or RV mass (i.e., a mass measurement ranges smaller than those provided by the above MR relationship) or eccentricity measurements for multiple planets, we generated 100 simulations with planetary masses (and eccentricities, where reported) assigned from within $1\sigma$ uncertainties of those measurements.
For planets in these systems without TTV/RV measurements, we assigned masses according to the same MR relationship as described above.
We sampled the same Rayleigh distribution in $e$ as above for planets without eccentricity constraints.
We assigned all planets inclinations from the Rayleigh distribution described above and randomized all other orbital angles.
For planets with multiple reported masses, we defaulted to the masses reported in the NASA Exoplanet Archive composite data table (references for our chosen masses are given in Table~\ref{tab:systems}). 
For two systems (Kepler-107, Kepler-11) that proved to be particularly prone to collisions, we ran simulations both with and without the reported eccentricity constraints.
We generated 4900 simulations representing the 47 observed systems and integrated them as described above.

\subsection{Summary of simulation outcomes}\label{ss:sim-summary}

Overall, we ran simulations of 21300 cases of Kepler-like planetary architectures. 
Of these, 2764 simulations ($\sim13$\%) ended with planet collisions occurring within $5\times10^9$ orbital periods of the innermost planet.
Of the 165 observed system architectures we consider, 60 experience at least one simulation that ends in a collision. 
When only considering the MR-relationship cases, 30\% of the observed systems (49/162) have at least one unstable simulation. 
In some contrast, 55\% of the system architectures for which we used reported planet masses (26/47) have at least one simulation that ends in collision.

\begin{figure}
    \centering
    \includegraphics[width=3.in]{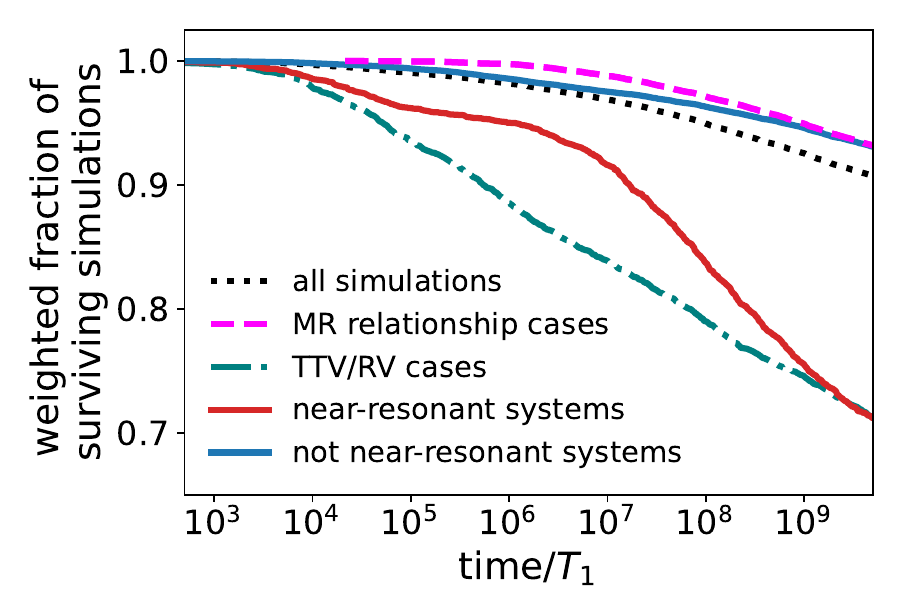}
    \caption{Fraction of surviving simulations vs time/$T_1$, where $T_1$ is the orbital period of the innermost planet in each individual simulation. 
    The simulations are weighted such that all observed systems in each simulation category contribute equally.}
    \label{f:instability-times}
\end{figure}

The distributions of instability times we found  in our simulations are summarized in Figure~\ref{f:instability-times}.
We show the results for the entire simulation set as well as the distributions for the MR relationship cases and the TTV/RV cases.
Additionally, we separate systems that contain near-resonant (or potentially actually resonant) planets and those that do not have planets in near-resonant configurations.
To classify the planets within systems as either potentially near or not near strong mean motion resonances (MMRs), we use analytical resonance width estimates from \cite{SSD1999} for a subset of low-order MMRs, including all internal and external first order MMRs from the 2:1 through the 8:7\footnote{
The full list of resonances and the code used to calculate their estimated widths is available on Github \url{https://github.com/katvolk/analytical-resonance-widths} \citep{Volk:2024} and described in \cite{Volk:2020}.}.
These analytical resonance width estimates are far from perfect, especially at very low eccentricities \citep[see, e.g.,][]{Malhotra:2023}, but they afford a quick way to estimate whether the observed period ratio of a pair of planets likely falls inside, near, or well outside a mutual MMR for some assumption about their masses. 
For each system we simulated, we include our assessment of their  resonant status for their initially low-eccentricity orbits in Table~\ref{tab:systems} as either likely non-resonant (no planets are near or in strong MMRs), likely resonant (at least one pair of planets is in or quite near a low order MMR), or ambiguous (apparently non-resonant, but modest eccentricity enhancement could bring a planet pair near a low order resonance).
For each curve in Figure~\ref{f:instability-times}, the simulations are weighted such that all observed systems contribute equally; for example, in the `all simulations' curve, systems that were simulated 200 times because they were included in both the MR relationship and TTV/RV cases were given weights of 50\% compared to systems that appeared in only one of those simulation sets and thus were simulated 100 times.

From Figure~\ref{f:instability-times}, it is clear that the near-resonant and TTV/RV cases have higher instability rates ($\sim30$\% of the simulations) than the non-resonant and MR relationship cases ($\sim7$\% of simulations). 
There is significant overlap between the observed systems that contribute to the curves for the near-resonant and TTV/RV cases as close proximity to strong MMRs increases the likelihood of detectable TTVs.
The near-resonant systems account for approximately one-third of all the collisions in our simulations. 
In addition to the higher rate of instabilities, the TTV/RV and near-resonant cases also have instabilities that occur on shorter timescales. 
This is consistent with some of these instabilities being directly related to the MMRs, which have shorter dynamical timescales than the secular dynamics that can also drive instabilities; this is discussed further in Section~\ref{sec:analysis}.
Overall, the rate of instabilities in the simulations is linear in $\log(t)$, which has has been found in previous studies of long term stability in ensembles of dynamical systems \citep[e.g.][]{Holman:1993,Chambers:2007,Volk:2015}.
We note that the shape of the near-resonant systems curve is distinct from the other curves; it displays a shallow slope similar to that of the non-resonant curve up to $10^7$ orbits before changing to a steeper slope. 
However, this could be the effect of a few observed system architectures as the number of observed systems contributing to the near-resonant curve is smaller than for any of the other curves. 
We discuss the distribution of key parameters for the unstable cases and the implications of these overall instability rates in more detail below.

\section{Stability as a function of different system parameters} \label{sec:analysis}

Here we examine the characteristics of all our simulated planetary architectures as well as individual simulations to identify the parameters most indicative of stability and instability.
We separate our entire suite of simulations into stable and unstable categories based on whether a collision occurred within the simulated $5\times10^9$ orbital periods of the innermost planet.
We also label system architectures (i.e., the combination of observed orbital periods and planetary radii in the MR-relationship cases or observed orbital periods and constrained masses in the TTV/RV cases) as either stable or unstable based on whether \textit{any} of the 100 simulations of that architecture experienced a collision.
For these two sets of our simulation suite, we then examine the similarities and differences in nearest-neighbor planet dynamical spacings, period ratios, and mass ratios.
We note that only $\sim$10\% of the collisions detected in our simulations occur between non-adjacent planets; in these cases, a planet's orbit becomes eccentric enough to cross the orbits of more than its nearest neighbor(s) and happens to collide with an initially non-neighboring planet.
We also examine stability as a function of the system-wide coefficient of variation in planet masses, which was recently suggested by \cite{Mishra:2023} as useful for characterizing planetary architectures.
Finally, we examine stability as a function of short-term secular evolution as characterized by the spectral fraction parameter, which was found by \cite{Volk:2020} to be predictive of long-term stability in planetary systems with four or more planets.
The details of all of the considered parameters are given below.

In all of our analyses below, individual simulations are weighted such that the sum of the number of all simulations of an observed adjacent planet pair is one; thus in our histograms there can be fractions ($<1$) of simulated planet pairs represented because for most observed pairs, each of the 100 simulations containing that pair has a weight of 0.01 (some observed pairs have lower weight if, for example, they have 200 simulations in the MR-relationship cases due to significantly differing radius measurements; see Sections~\ref{sec:MR-ic} and \ref{sec:non-MR-ic}).
We examine stability as a function of our dynamical parameters in two ways: 1) examining the distribution of parameters when observed architectures are labeled as entirely stable (i.e., none of the 100 simulations of that architecture experienced a collision) vs architectures with at least one collision, and 2) on a simulation-by-simulation basis, separating simulations with and without collisions. 
We also examine separately our simulations that use a mass-radius relationship for assigning planet masses and those that use RV or TTV mass constraints.

\begin{figure}
    \centering
    \includegraphics[width=3.in]{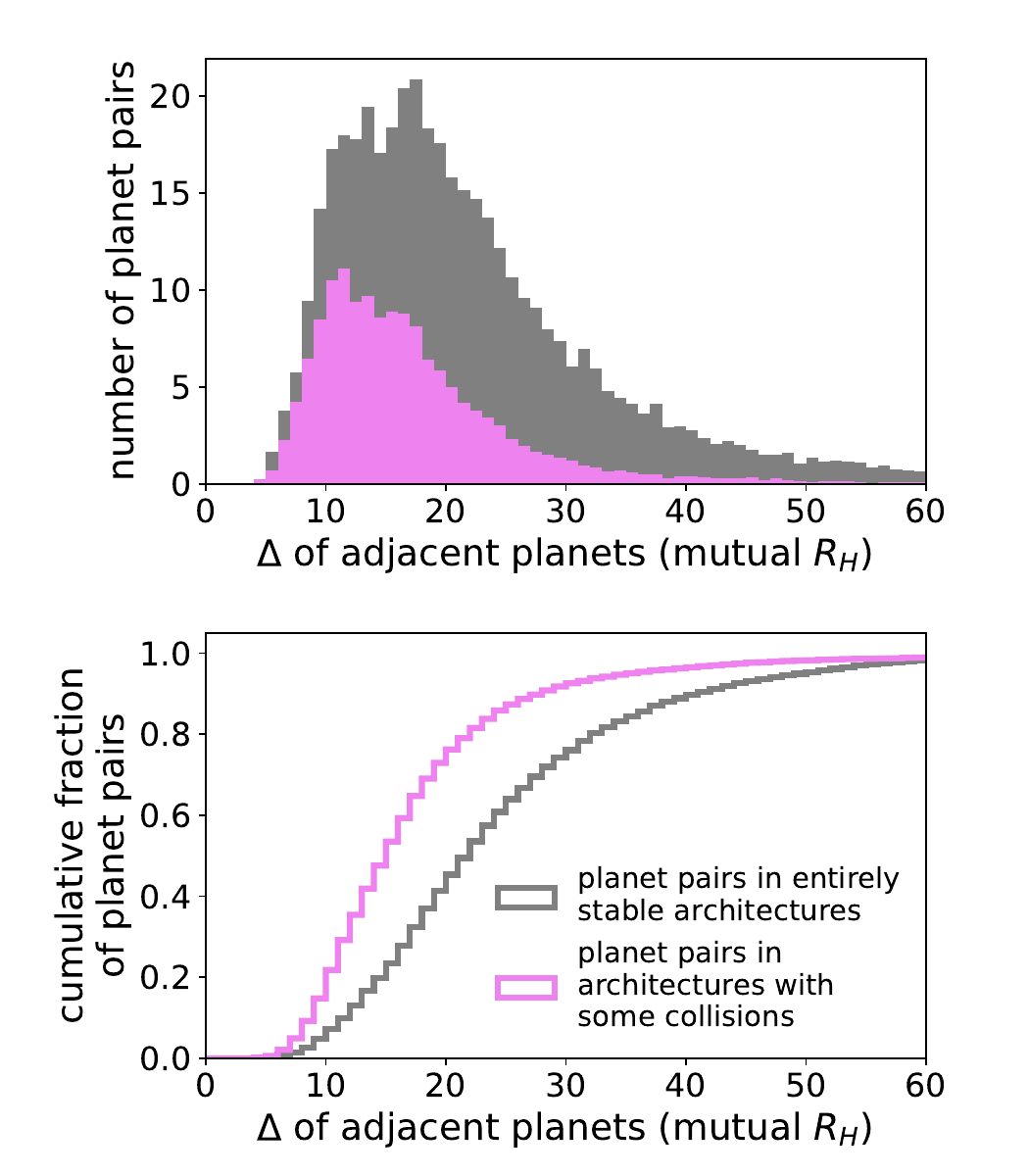}
    \caption{MR-relationship cases: Distribution of mutual hill radius separation ($\Delta$) between neighboring planets in 162 observed planetary system architectures with 3 or more planets (top: stacked histogram with simulations weighted to sum to the observed number of planet pairs; bottom: normalized cumulative distributions). 
    In grey are pairs of planets in system architectures where no collisions occurred within $5\times10^9$ orbits of the shortest period planet across 100 instances of that architecture (initial conditions described in Section~\ref{sec:MR-ic}). In pink are planet pairs in system architectures for which at least one of the 100 instances experienced a collision within $5\times10^9$ inner planet orbits.  }
    \label{fig:delta_MR}
\end{figure}

\begin{figure}
    \centering
    \includegraphics[width=3in]{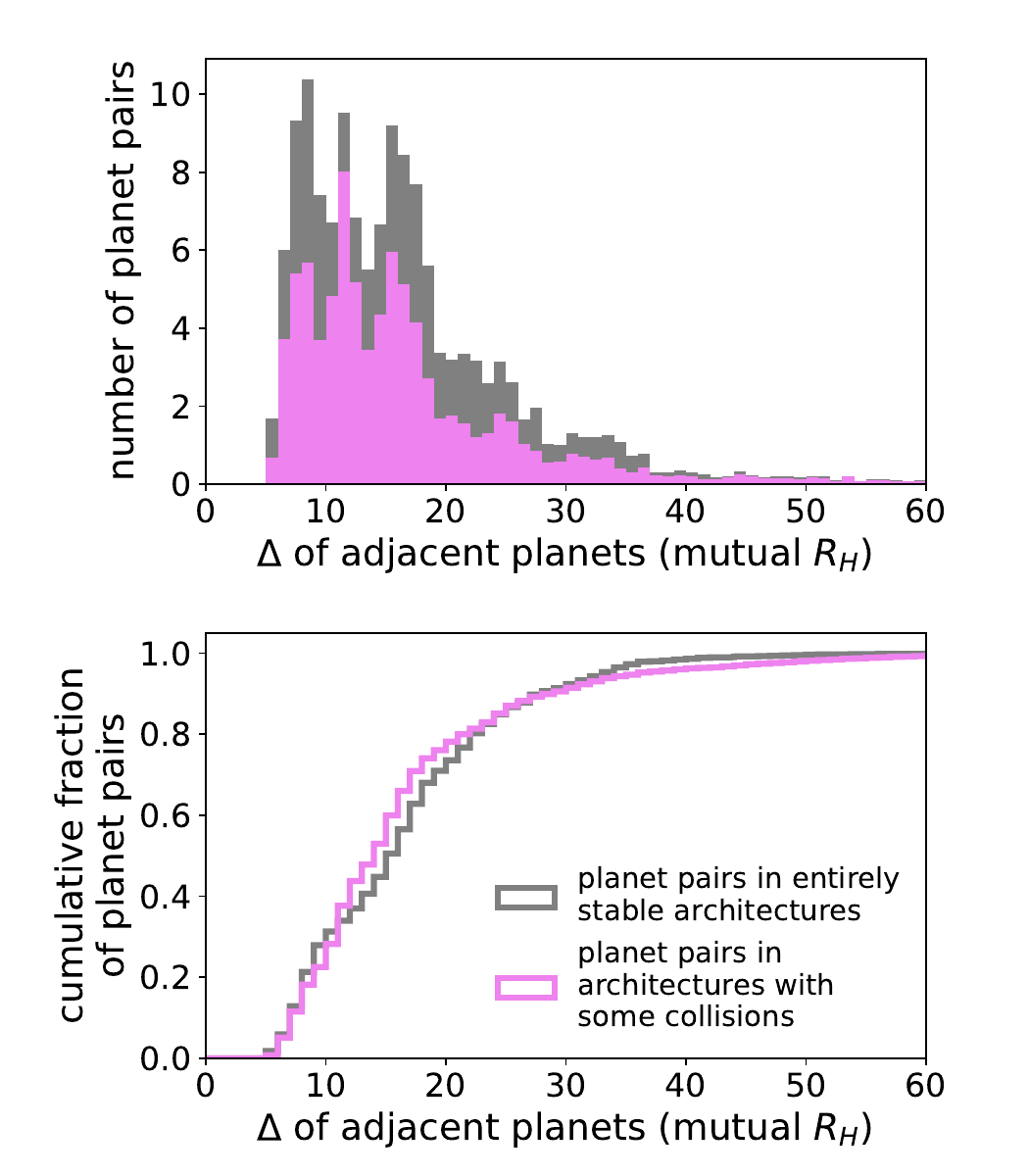}
    \caption{TTV/RV cases: Distribution of mutual hill radius separation ($\Delta$) between neighboring planets in 47 observed planetary system architectures with 3 or more planets and TTV or RV mass constraints on at least one planet (top: stacked histogram weighted to the observed number of planet pairs; bottom: normalized cumulative distributions). 
    In grey are pairs of planets in system architectures where no collisions occurred within $5\times10^9$ orbits of the shortest period planet across 100 instances of that architecture (initial conditions described in Section~\ref{sec:non-MR-ic}). 
    In pink are planet pairs in system architectures for which at least one of the 100 instances experienced a collision within $5\times10^9$ inner planet orbits.  }
    \label{fig:delta_nonMR}
\end{figure}

We begin with examining the dynamical separations, $\Delta$, of neighboring planets with semimajor axes $a_1$ and $a_2$ and masses $m_1$ and $m_2$ in units of their mutual hill radius $R_{mH}$:
\begin{equation}\label{eq:delta}
\Delta = \frac{|a_1 - a_2|}{R_{mH}} = \frac{2|a_1 - a_2|}{a_1+a_2}\left[\frac{3M_*}{m_1+m_2}\right]^{1/3}
\end{equation}
where $M_*$ is the mass of the planets' host star.
Figure~\ref{fig:delta_MR} shows the distribution of $\Delta$ in our MR-relationship cases; Figure~\ref{fig:delta_nonMR} shows the same distributions for our TTV/RV cases.
In Figures~\ref{fig:delta_MR} and~\ref{fig:delta_nonMR}, system architectures in which no collisions occurred (i.e., all 100 simulations reached $5\times10^9$ orbits of the innermost planet) are shown in grey while system architectures with at least one collision (out of the 100 simulations) are shown in pink; note that these histograms are stacked.

Figure~\ref{fig:delta_MR} shows that a large fraction of the system architectures with planet pairs at $\Delta\lesssim10$ are prone to instabilities; this is consistent with previous investigations  \citep[e.g.,][]{Gladman:1993,Chambers:1996,Smith:2009,Funk:2010,Malhotra:2015,Pu:2015,Obertas:2017,Lissauer:2021}. 
The peak in the pink histogram (architectures with unstable cases) occurs at a smaller value of $\Delta$ than in the grey histogram (stable architectures); the bottom panel of Figure~\ref{fig:delta_MR} confirms that the two distributions are distinct.
There are two additional notable features in the $\Delta$ distribution of the unstable architectures:
(i) the peak is relatively flat in the range $\sim12-18$, and (ii) the distribution extends to fairly large values (consistent with findings in \citealt{Volk:2015}), with $\gtrsim10$\% of planet pairs at separations larger than 25 mutual hill radii.
We note that the $\Delta$ distributions of entirely stable vs unstable architectures for the systems with TTV/RV mass measurements (Figure~\ref{fig:delta_nonMR}) are less distinct from each other than in the MR relationship cases (Figure~\ref{fig:delta_MR}), but the unstable architectures span a similar range of $\Delta$ as in the MR relationship simulations.
It is also notable that a larger fraction of the stable TTV/RV system architectures have planet pairs with $\Delta<10$ ($\sim30$\% of planet pairs in the TTV/RV cases compared to $<5$\% in the MR cases).

\begin{figure}
    \centering
    \includegraphics[width=3in]{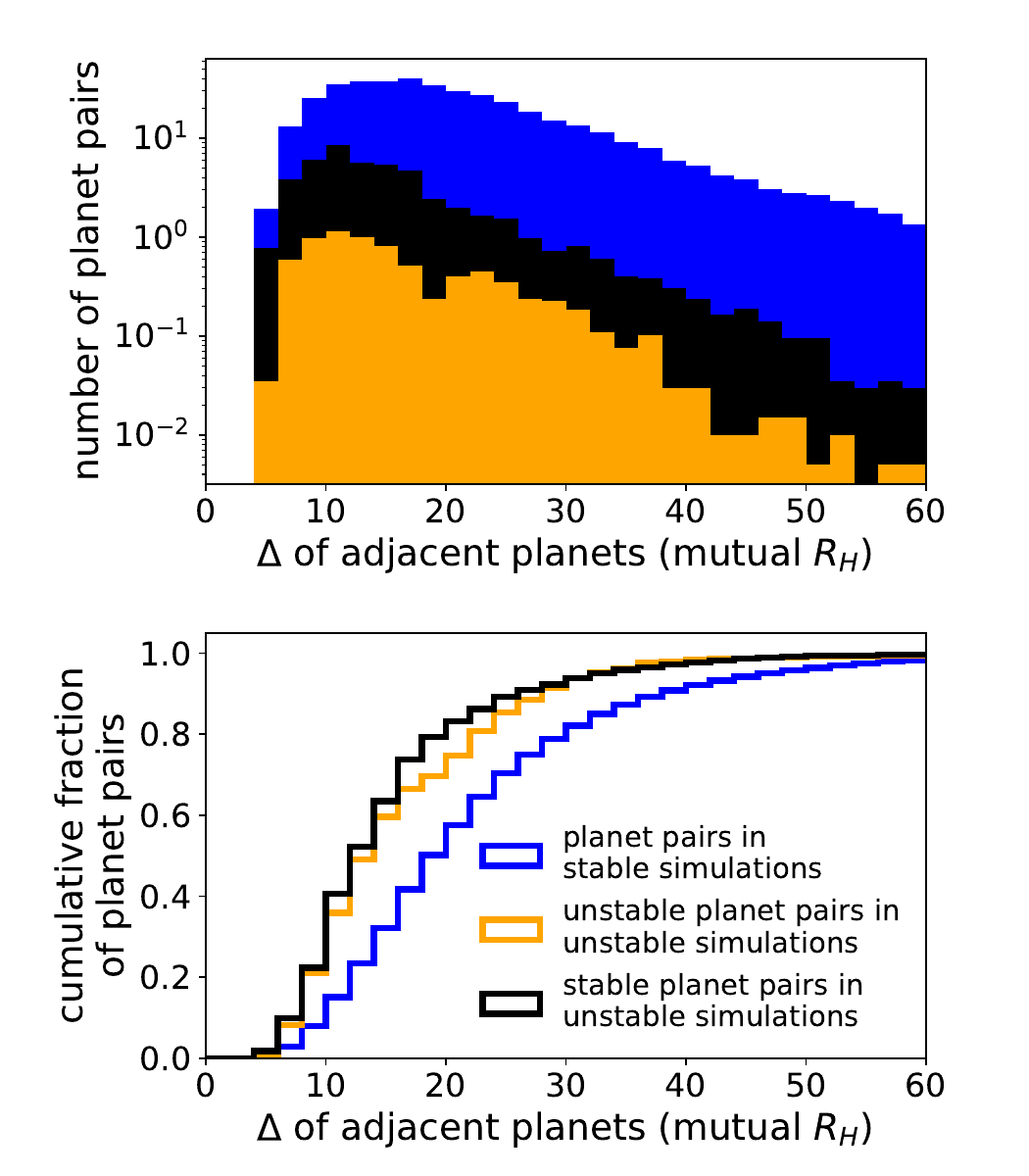} 
    \caption{All simulations: Distribution of mutual hill radius separation ($\Delta$) between neighboring planets (top: stacked histograms that sum to the observed number of planet pairs; bottom: normalized cumulative distributions). 
    Note the log scale in the top panel used to better highlight the distribution of the colliding pairs.
    Colliding planet pairs are shown in orange with the other adjacent planet pairs in those simulations shown in black. 
    Adjacent planet pairs in simulations with no collisions are shown in blue. Note that a few percent of the colliding pairs of planets were not adjacent at the start of the simulations.
    }
    \label{fig:delta_pairs}
\end{figure}

Figure~\ref{fig:delta_pairs} shows the $\Delta$ distribution of the actually colliding pairs of planets compared to both the non-colliding pairs in unstable simulations and planet pairs in stable simulations. 
This figure combines all simulations regardless of the source of the planet masses, and we use a log scale on the histogram to better highlight the colliding planet pairs.
We see that collisions do indeed occur between planets at initially quite large dynamical separations, with $\sim30$\% of collisions occurring between planets initially separated by $\Delta>20$.

We note that $\Delta$, which scales with the planet masses to the one-third power (Equation~\ref{eq:delta}), is not the only possible formulation for dynamical separations. 
Resonance overlap considerations yield a dynamical separation parameter that instead scales with planet mass to the one-fourth power \citep[e.g.][]{Hadden:2018,Tamayo:2021}.
For the simulations presented here, both variations of the dynamical separation parameter yield very similar results.

\begin{figure}
    \centering
        \includegraphics[width=3.in]{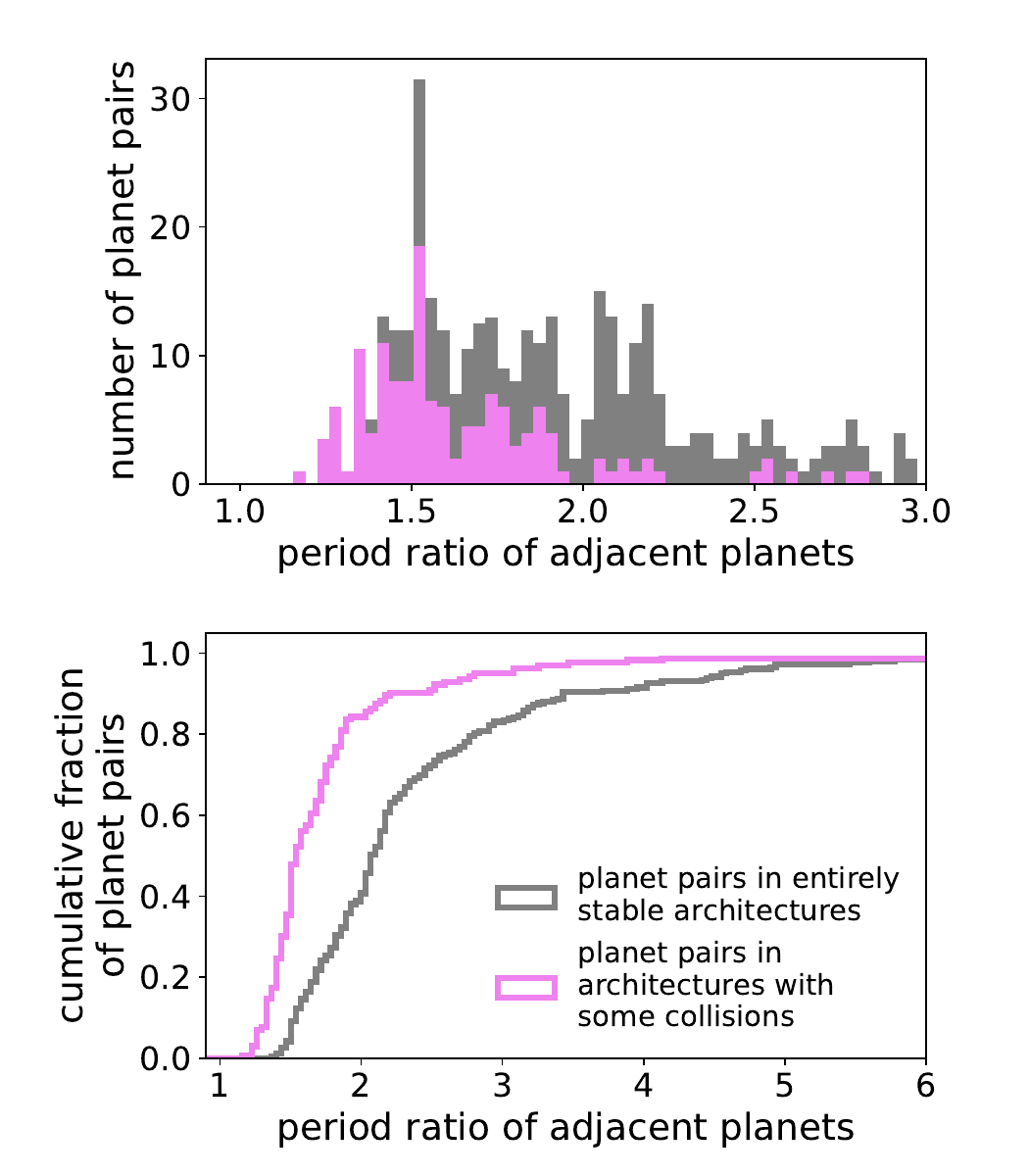}
    \caption{MR-relationship cases: Distribution of period ratios between neighboring planets in 162 observed planetary system architectures with 3 or more planets (top: stacked histogram that sums to the observed number of planet pairs; bottom: normalized cumulative distributions).  
    In grey are pairs of planets in system architectures where no collisions occurred within $5\times10^9$ orbits of the shortest period planet across 100 instances of that architecture (initial conditions described in Section~\ref{sec:MR-ic}). 
    In pink are period ratios in system architectures for which at least one of the 100 instances experienced a collision within $5\times10^9$ inner planet orbits.  }
    \label{fig:pr_MR}
\end{figure}

\begin{figure}
    \centering
        \includegraphics[width=3.in]{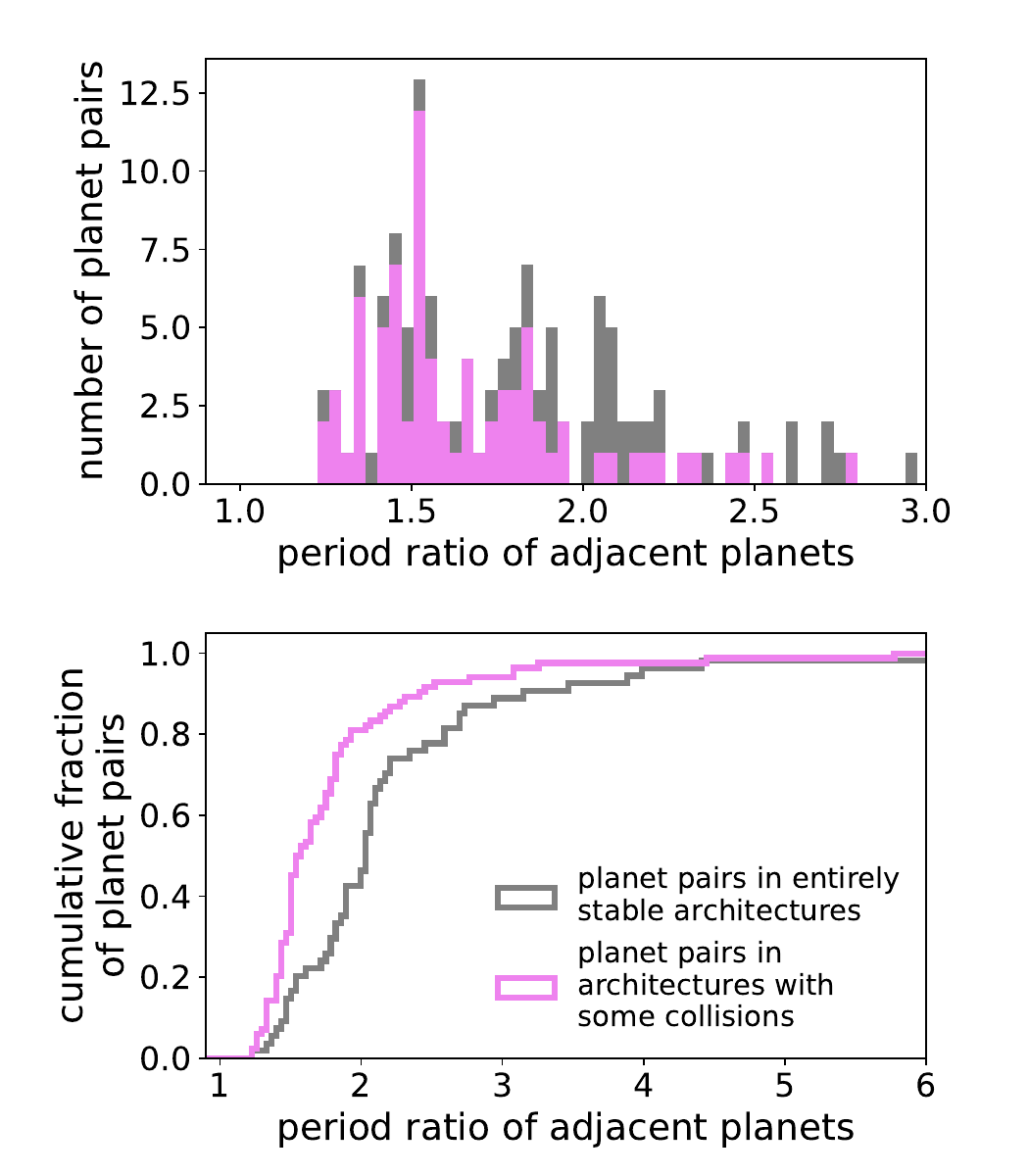}  \\
    \caption{TTV/RV cases: Distribution of period ratios between neighboring planets in 47 observed planetary system architectures with 3 or more planets and TTV or RV mass constraints on at least one planet (top: stacked histogram that sums to the observed number of planet pairs; bottom: normalized cumulative distributions). 
    In grey are pairs of planets in system architectures where no collisions occurred within $5\times10^9$ orbits of the shortest period planet across 100 instances of that architecture (initial conditions described in Section~\ref{sec:non-MR-ic}). 
    In pink are planet pairs in system architectures for which at least one of the 100 instances experienced a collision within $5\times10^9$ inner planet orbits.    }
    \label{fig:pr_nonMR}
\end{figure}

We now examine the period ratio distributions of neighboring planets in the systems simulated; note that we define the period ratio to be larger than one (i.e., for each pair we take the period of the outer planet divided by the inner planet).
Figure~\ref{fig:pr_MR} shows this distribution in our simulations based on MR relationship cases and Figure~\ref{fig:pr_nonMR} shows the same distributions for TTV/RV cases.
It is immediately notable that the period ratio distribution is much less smooth than the $\Delta$ distributions. 
This is partly due to the nature of the dataset on which we base our sample.
Due to the large uncertainty in each planet's mass, the $\Delta$ value for each observed planet pair is, in effect, spread out over a wide range in the 100 simulation instances; in both the MR and TTV/RV cases, these uncertainties smooth out the contributions to the distribution for each observed planet pair. 
In contrast, each planet's orbital period is much better constrained by the observations and has a very small uncertainty, so every instance of each simulated planet pair has a nearly identical period ratio.
However, there is also an expectation on grounds of dynamical stability that the $\Delta$ distribution is fairly smooth \citep[e.g.][]{Dietrich:2023} whereas the period ratio distribution has strong variability near first order resonances.
In particular, there are notable features in the period ratio distribution near the strongest mean motion resonances (MMRs), e.g., the peak at the 3:2 MMR and the trough near the 2:1 MMR, as discussed in previous studies of the Kepler sample \citep[e.g.,][]{Fabrycky:2014,Steffen:2015}.

\begin{figure}
    \centering
    \includegraphics[width=3.in]{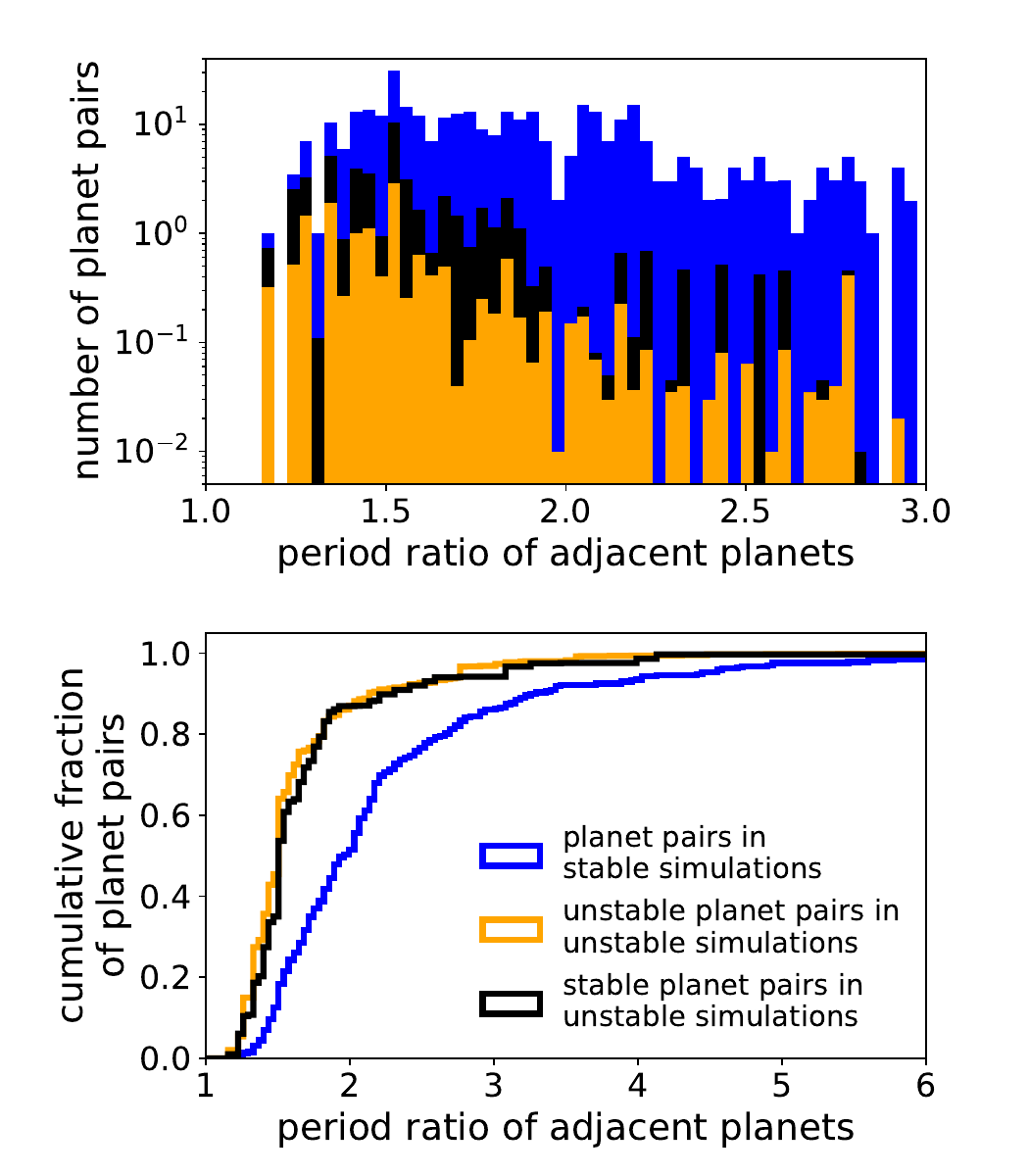} 
    \caption{All simulations: Distribution of period ratios between neighboring planets (top: stacked histogram that sums to the observed number of planet pairs; bottom: normalized cumulative distributions).
    Note the log scale in the top panel used to better highlight the distribution of the colliding pairs.
    Colliding planet pairs are shown in orange with the other adjacent planet pairs in those simulations shown in black. Planet pairs in simulations with no collisions are shown in blue. Note that a few percent of the colliding pairs of planets were not adjacent at the start of the simulations.}
    \label{fig:pr_pairs}
\end{figure}

A very striking result of our suite of simulations is that $\sim90\%$ of planet pairs in system architectures that are prone to collisions are at period ratios $\lesssim2$ (see the bottom panels of Figures~\ref{fig:pr_MR} and \ref{fig:pr_nonMR}). 
This can also be seen in Figure~\ref{fig:pr_pairs}, which shows the distribution of period ratios for colliding pairs of planets across both sets of our simulations compared to the period ratios of other planet pairs in those simulations as well as planet pairs in stable simulations.
Some of this tendency can likely be explained by proximity; planets with smaller period ratios are on orbits that are spatially closer together than those with larger period ratios, and thus require only modest eccentricities to achieve crossing orbits.
There is also a higher density of low-order mean motion resonances at period ratios less than two, which can enhance dynamical chaos and eccentricity excitation.
We will discuss the implications of this period ratio trend below as well as in Section~\ref{sec:discussion}.

\begin{figure}
    \centering
        \includegraphics[width=3.in]{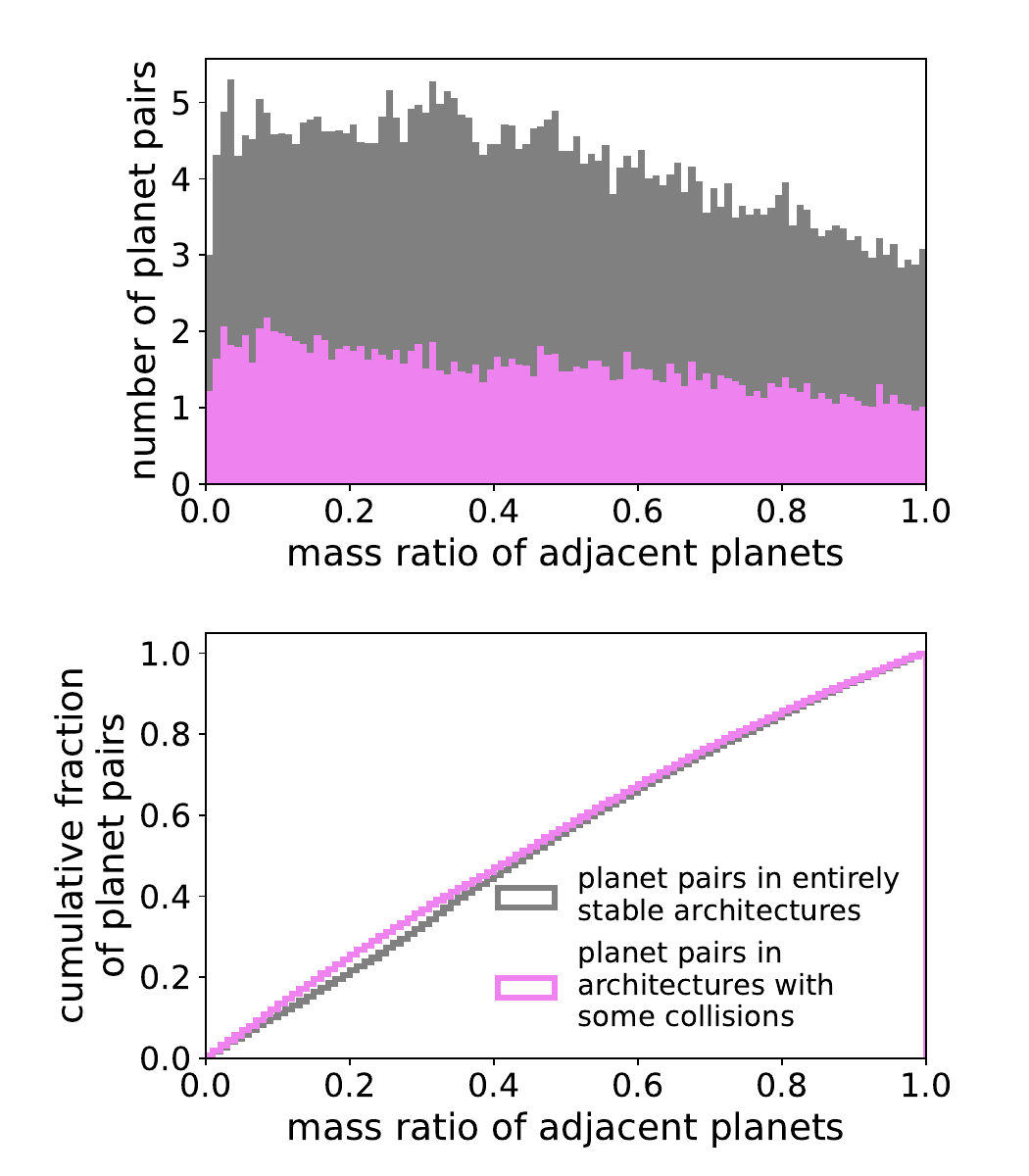}
    \caption{MR-relationship cases: Distribution of mass ratios between neighboring planets in 162 observed planetary system architectures with 3 or more planets (top: stacked histogram that sums to the observed number of planet pairs; bottom: normalized cumulative distributions).  
    In grey are pairs of planets in system architectures where no collisions occurred within $5\times10^9$ orbits of the shortest period planet across 100 instances of that architecture (initial conditions described in Section~\ref{sec:MR-ic}). 
    In pink are period ratios in system architectures for which at least one of the 100 instances experienced a collision within $5\times10^9$ inner planet orbits.    }
    \label{fig:mr_MR}
\end{figure}

\begin{figure}
    \centering
        \includegraphics[width=3.in]{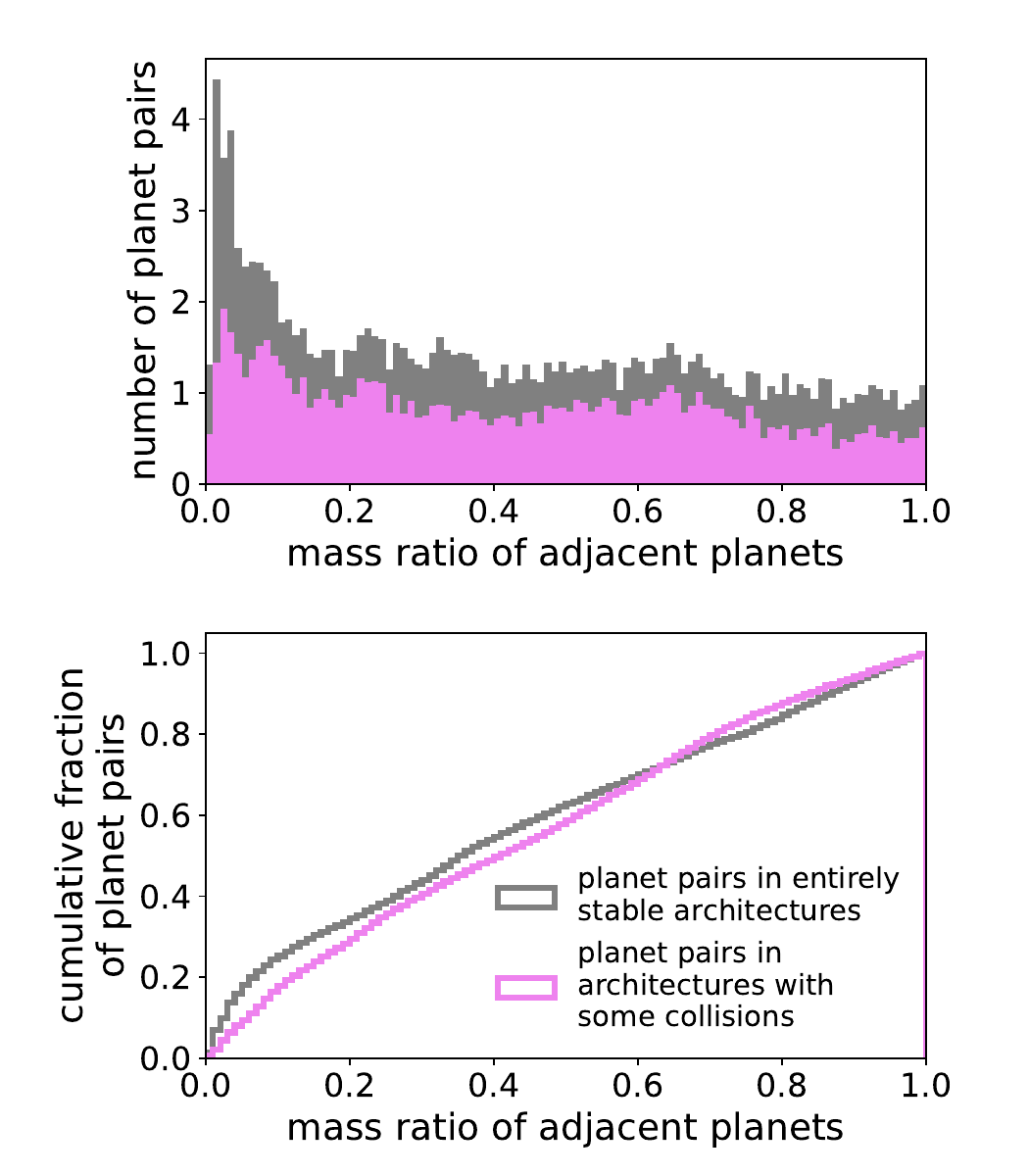}
    \caption{TTV/RV cases: Distribution of mass ratios between neighboring planets in 47 observed planetary system architectures with 3 or more planets and TTV or RV mass constraints on at least one planet (top: stacked histogram that sums to the observed number of planet pairs; bottom: normalized cumulative distributions). 
    In grey are pairs of planets in system architectures where no collisions occurred within $5\times10^9$ orbits of the shortest period planet across 100 instances of that architecture (initial conditions described in Section~\ref{sec:non-MR-ic}). 
    In pink are planet pairs in system architectures for which at least one of the 100 instances experienced a collision within $5\times10^9$ inner planet orbits.     }
    \label{fig:mr_nonMR}
\end{figure}

\begin{figure}
    \centering
     \includegraphics[width=3.in]{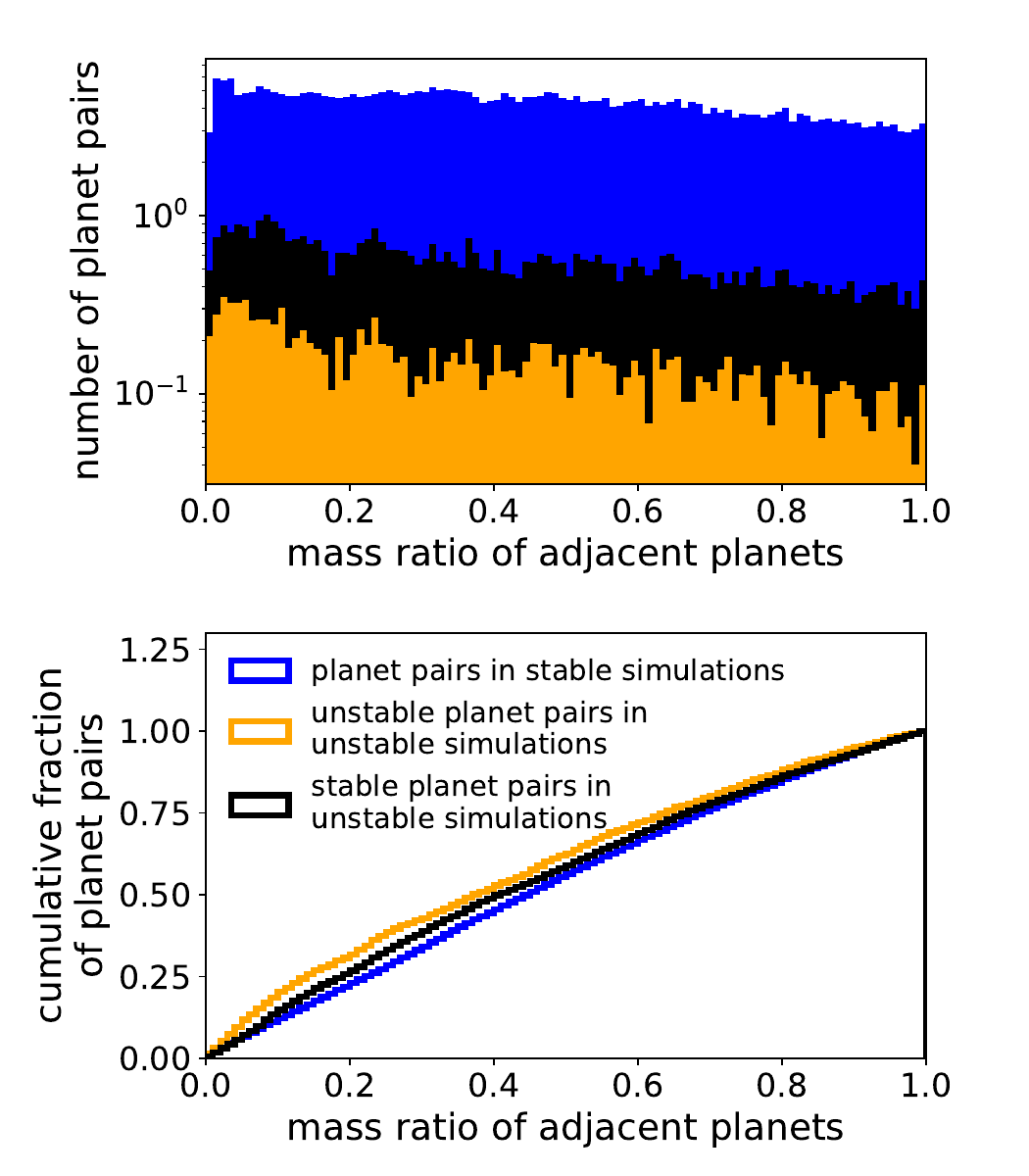} 
    \caption{All simulations: Distribution of mass ratios between neighboring planets (top: stacked histogram weighted to the observed number of planet pairs; bottom: normalized cumulative distributions). 
    Note the log scale in the top panel used to better highlight the distribution of the colliding pairs.
    Colliding planet pairs are shown in orange with the other adjacent planet pairs in those simulations shown in black. Planet pairs in simulations with no collisions are shown in blue. Note that a few percent of the colliding pairs of planets were not adjacent at the start of the simulations.}
    \label{fig:mr_pairs}
\end{figure}

We also examine the mass ratios of neighboring planets in our systems; note that we define the planet mass ratio to be the smaller of $m_{inner}/m_{outer}$ or $m_{outer}/m_{inner}$, so it is always less than one.
Figures~\ref{fig:mr_MR} and~\ref{fig:mr_nonMR} show the distribution of mass ratios for neighboring planets in the MR relationship cases and the TTV/RV cases, respectively.
There are no strong apparent trends in these plots to differentiate the stable architectures from those that are prone to instabilities. 
It is notable, however, that the TTV/RV cases have a higher abundance of neighboring planets with smaller mass ratios (more unequal masses).
Figure~\ref{fig:mr_pairs} shows the distribution of mass ratios for the colliding pairs of planets in all unstable simulations compared to the non-colliding planets in those simulations and the planet pairs in the stable simulations. 
We note a weak trend that the unstable simulations have slightly smaller mass ratios than the stable simulations, and that the colliding planet pairs in the unstable simulations have slightly smaller mass ratios than the stable pairs in those simulations.

\begin{figure}
    \centering
    \includegraphics[width=3.in]{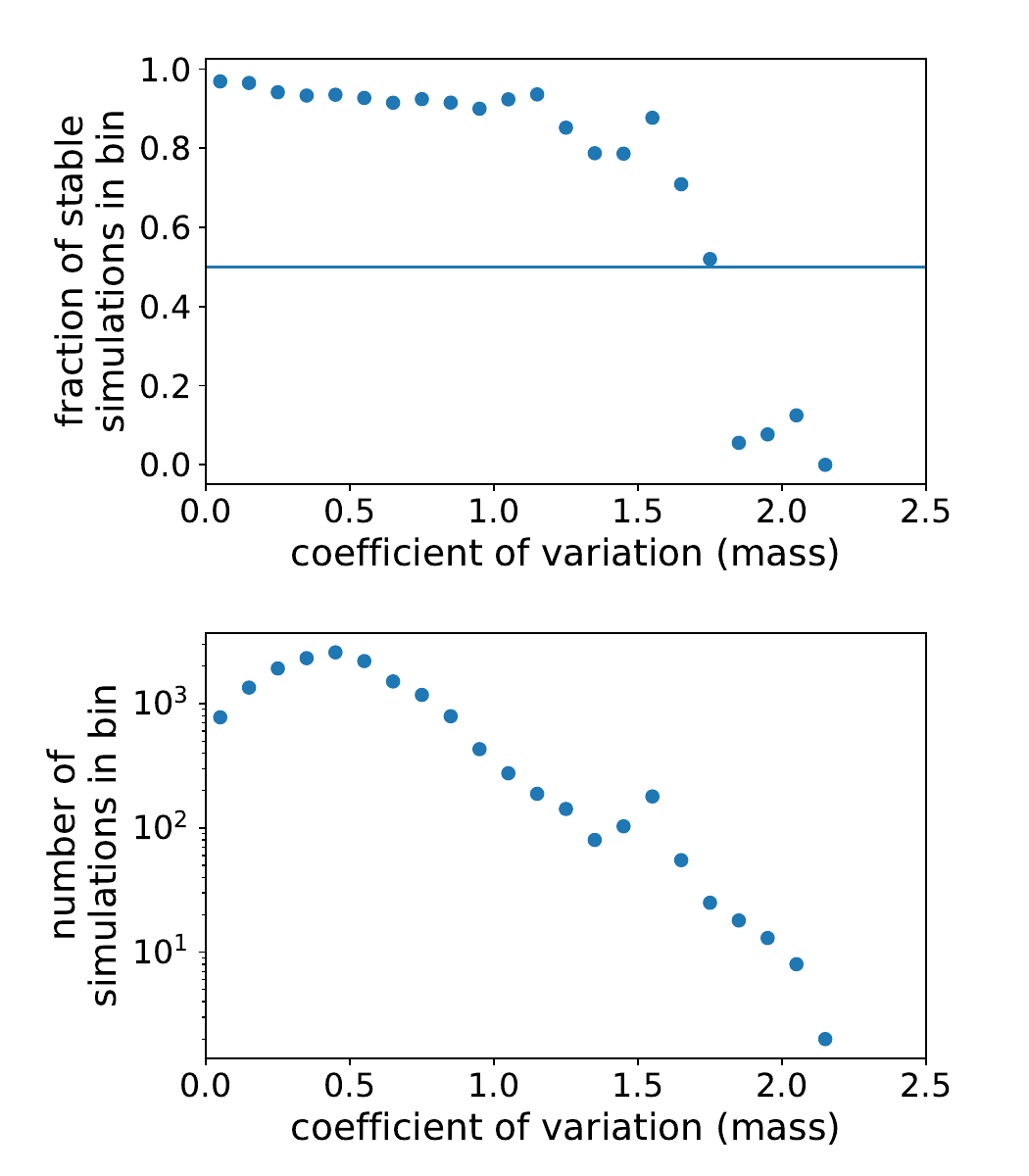}
    \caption{MR-relationship cases: Fraction of stable simulations as a function of the coefficient of variation for the masses of the planets in each simulation (top) and number of simulations per bin in that coefficient (bottom).}
    \label{fig:mass_crcv}
\end{figure}

Following \cite{Mishra:2023}, we also examine the mass distributions in our individual simulations using the coefficient of variation in planet masses; this coefficient, $C_{v}$ is defined as the standard deviation of the planet masses in the system divided by the average planet mass.
Figure~\ref{fig:mass_crcv} shows this characterization of system-wide planet masses for all of our MR relationship simulations. 
The vast majority of our simulated systems have reasonably small coefficients of variation for the planet masses ($C_v\lesssim1$, meaning the planet masses vary by less than the average planet mass).
Across this well-sampled range, the fraction of stable simulations decreases roughly linearly with increasing $C_v$, from $\sim98\%$ in the smallest $C_v$ bin to $\sim90\%$ at $C_v\approx1$.
The trends at larger $C_v$ values are noisier due to the smaller number of simulated systems, but the stable fraction drops off for $C_v\gtrsim1$ and is very low at the largest $C_v$ values.
This trend in $C_v$ is consistent with the weak trend noted above that unstable systems are more common for more unequal planetary mass ratios.
The modest $\sim10$\% decrease in the stable fraction may contribute to the high abundance of systems with nearly equal size planets (the so-called `peas-in-a-pod' trend, discussed in, e.g. \citealt{Weiss:2020,Weiss:2023}). 
We discuss this further in Section~\ref{sec:discussion}.

Lastly, we examine stability as a function of the secular dynamics of each simulated system.
Briefly, classical linear secular theory describes the orbit-averaged perturbations of the planets on each other, which cause their orbital eccentricities and inclinations to vary over time in a quasi-periodic way with a superposition of several secular modes. 
Beyond the linear secular approximation, the secular evolution of multi-planet systems is more complex than the simple superposition of a few secular modes.
We characterize the secular dynamics of our simulated planetary systems using the spectral fraction parameter from \cite{Volk:2020}, which is a modification of the spectral number described by \citet{Michtchenko:2002} for asteroid dynamics. 
To calculate a spectral fraction, we perform a short integration ($5\times10^6$ orbits of the innermost planet) of each of our simulated systems with high-cadence output of the planets' orbital parameters (3000 outputs evenly spaced in time). 
We then perform a Fast Fourier Transform (FFT) of the orbital parameter of interest for a planet and identify the frequency with the highest power (the square of the amplitude) in the FFT.
We determine how many of the frequencies sampled in the FFT have a power $\ge5$\% of that maximum power, and this number divided by the total number of frequencies in the FFT 
is then the spectral fraction for that orbital parameter for that planet.

We considered the spectral fraction of many dynamically meaningful parameters, and we determined empirically that the results for the angular momentum deficit, {\sc{amd}}, of each planet were the most relevant to predicting long-term stability.
The {\sc{amd}} of an individual planet in a system is given by:
\begin{equation}
\text{\sc{amd}}_j=m_j\sqrt{Gm_*a_j}(1-\sqrt{1-e_j^2}\cos i_j),
\end{equation}
where the index $j$ indicates the $j$-th planet's parameters.
The {\sc{amd}} spectral fraction captures the same stability trends we see in the $e$ and $i$ spectral fractions, allowing us to rely on just one parameter rather than two; we note that the semimajor axis spectral fraction is not strongly correlated with stability in our simulation set.
Following \cite{Volk:2020}, we choose the largest spectral fraction of the {\sc{amd}} of any planet in each system to characterize the whole system.
We note that some of our simulations ended in a collision within the short $5\times10^6$ orbit simulations used for the spectral fractions. 
We only calculate spectral fractions for simulations that survived at least $10^6$ orbits (the minimum time to compute the FFT of the {\sc{amd}} over secular timescales), so our analysis here excludes shorter-lived systems.
The limited length of our spectral fraction integrations does limit how well the low-frequency evolution of the planets is captured by the spectral fraction parameter. 
The integration length of $5\times10^6$ orbits was chosen with both efficiency and expected secular timescales in mind.
However there may be individual systems, e.g. ones with outer planets at orbital periods particularly larger than the innermost planet, where longer integration timescales are necessary to adequately determine the nature of their secular evolution.
The general usefulness of the spectral fraction calculated from our chosen integration length (see below) indicates that timescales important to instabilities are reasonably well-captured for the systems we investigate here, but this limitation should be considered when applying the spectral fraction calculation in future investigations.

\begin{figure}
    \centering
    \includegraphics[width=3.in]{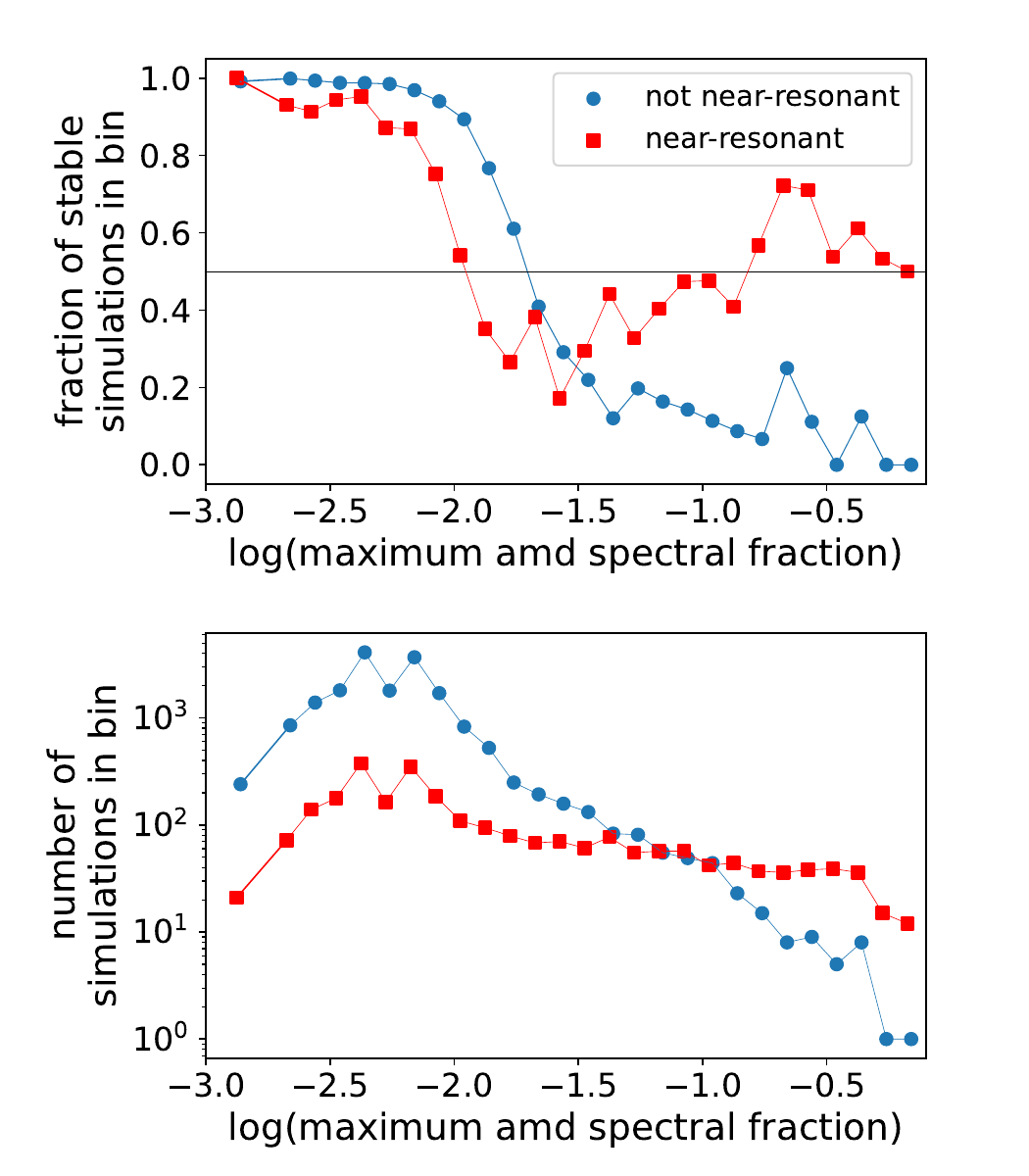}
    \caption{All simulations: Fraction of stable simulations as a function of the largest {\sc{amd}} spectral fraction for any planet in that simulation (top) and number of simulations per spectral fraction bin (bottom).
    Planetary systems without planet pairs starting obviously near strong mean motion resonances (systems with MMRs labeled as 0 or 2 in Table~\ref{tab:systems}) are shown as blue dots, while those that are likely in or near resonant are shown as red squares (systems with MMRs labeled as 1 in Table~\ref{tab:systems}). 
    These plots include all simulations performed (whether the planets masses were assigned from the mass-radius relationship or from TTV/RV measurements) that survived for at least $10^6$ orbits, the minimum timespan to perform a useful FFT.
    \label{fig:amd-sf}
    }
\end{figure}

\begin{figure}
    \centering
    \includegraphics[width=3.in]{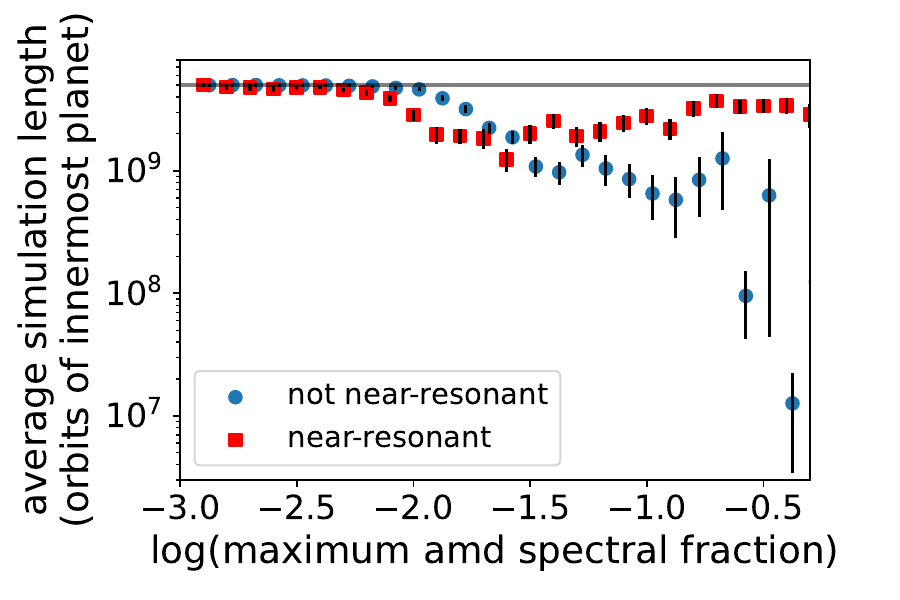}
    \caption{All simulations: Average simulation stopping time (in units of inner-most planet orbital periods) vs maximum {\sc{amd}} spectral fraction for systems not obviously near mean motion resonances (blue dots) and those that are likely resonant or near-resonant (red squares).
    The standard deviation of the simulation lengths in each spectral fraction bin are shown with black error bars, except for the right-most three bins, which only have one simulation per bin.
    Note that the two sets of data points are artificially offset on the horizontal axis to better differentiate the points.
    Simulations ran until $5\times10^9$ orbits of the innermost planet (grey horizontal line) or until two planets collided. }
    \label{fig:time-vs-sf}
\end{figure}

Figure~\ref{fig:amd-sf} shows the stable fraction of our simulations versus the maximum {\sc{amd}} spectral fraction for systems that appear not to be near any strong mean motion resonances and systems that are potentially in or near strong resonances (see Section~\ref{ss:sim-summary}).
It is useful to separate systems in this way because our definition of the spectral fraction is motivated by considering secular rather than resonant dynamics.
In both the near-resonant and not near-resonant simulations, the highest stable fractions are at the lowest values of the {\sc{amd}} spectral fractions.
The trend of decreasing stability with increasing spectral fraction is clear in the non-resonant systems; this trend is also clear in the near-resonant systems at small spectral fractions, though stability flattens out and becomes noisier at {\sc{amd}} spectral fractions larger than $\sim0.01$ (Figure~\ref{fig:amd-sf}). 
This supports the idea that secular chaos is driving the onset of instability in the non-resonant systems; secular chaos can play a role in the resonant and near-resonant systems, but the spectral fraction is not as clearly predictive of stability or instability as in the non-resonant cases.
Figure~\ref{fig:time-vs-sf} shows the average simulation length for the non-resonant systems as a function of the maximum {\sc{amd}} spectral fraction. 
Over the well-sampled range of spectral fractions (values $\lesssim0.1$), there is a clear trend toward shorter stability times with increasing spectral fraction. 
We discuss the implications of this further in Section~\ref{sec:discussion}.

\section{Notable results for individual Kepler and K2 systems} \label{sec:individual-systems}

While none of our simulations are meant to perfectly reproduce the observed {\it Kepler} or {\it K2} systems they are based on, they still offer insights into the likely dynamics and properties of the real systems. 
Here we highlight interesting results for a subset of our systems and compare them to other studies that have examined these systems in detail. 
We discuss three systems that are potentially in resonant chains (Kepler-223, Kepler-60, and Kepler-80) and the role those resonances play in their stability. Next, we discuss the K2-138 system, comparing our simulations of its system architecture before and after confirmation of a sixth planet. Finally, we discuss a number of systems with particularly unstable architectures: Kepler-23, Kepler-402, Kepler-444, Kepler-1542, and K2-384.

\smallskip
\noindent\textit{Kepler-223}: The four planets in Kepler-223 are observed to be in a resonant chain \citep{Mills:2016} with period ratios of 4/3 (e/d), 3/2 (d/c), and 4/3 (c/b). 
When adopting the nominal TTV-derived masses and eccentricities for all four planets (Table 1 of \citealt{Mills:2016}), 80\% of our simulations are unstable within five billion orbits of planet b (101 Myr). 
When adopting MR derived masses for the inner two planets (the outer two are too large for the assumed MR relationship), and eccentricities chosen from a Rayleigh distribution of width $\sigma_e=0.02$ for all planets, only 21\% of the simulations become unstable on the same timescale.
There is not a strong correlation between stability and the assumed masses of the inner two planets (our assumed MR distribution overlaps the TTV mass constraints), but the TTV-derived eccentricities for planets b and c are large enough to place them initially on orbits that are not  far from crossing orbits, hence close to instability unless the planets are deep in resonance libration.
It thus seems likely that libration in resonance is important to maintaining long-term stability at larger eccentricities.
Interestingly, however, we find a similar rate of stability on 10 Myr timescales for our simulations based on the TTV-constrained masses as \cite{Mills:2016} (25\% of our simulations vs 30\% of theirs) despite our simulations randomizing all the planets' orbital angles; the \cite{Mills:2016} simulation initial conditions were based on photodynamical modeling, matching all dynamical parameters to the observed transits.
A statistically equivalent percentage of both sets of the 10 Myr stable simulations are also stable to 100 Myr; from their 10 Myr stable simulations they randomly selected to extend 25 of them, 16 of which were stable compared to 20 of our 25 simulations.

\cite{Mills:2016} also ran a set of simulations based on orbital parameters that ensure libration of the Laplace angles associated with the resonant chain, finding 100\% of their resulting 100 Myr simulations were stable; they thus argue that libration in the resonant chain is favorable to long-term stability.
This simulation set also had smaller planetary eccentricities than their nominal TTV solution.
It is interesting to note that in our low-eccentricity, MR mass simulation set, with no constraints on planetary angles, 79\% of the simulations are still stable at 100 Myr. 
Our MR simulations have slightly lower planetary eccentricities than the \cite{Mills:2016} resonant solution set, particularly for planet b (see their Extended Data Table 3), but no constraints on resonant angles.
It is indeed likely, as \cite{Mills:2016} argue, that libration in the resonant chain promotes long-term stability given that our stability rate is smaller than theirs. 
However, our simulations show that smaller  eccentricities can account for some portion of enhanced stability, so libration within the resonant chain might not be strictly required for long-term stability.

\smallskip
\noindent\textit{Kepler-60}: The three observed planets in Kepler-60 are potentially in a resonant chain, with period ratios of 4/3 (d/c) and 5/4 (c/b). 
When adopting MR masses for the planets, 61\% of our simulations are unstable over the 97 Myr simulations.
However, none of our simulations are unstable when we adopt the TTV planet masses from \cite{Jontof-Hutter:2016} despite not imposing any constraints on the relative orbital phases of the three planets in the system.
\cite{Jontof-Hutter:2016} did complete fits to the transit data and found that both resonant and non-resonant solutions are consistent with the data.
\cite{Gozdziewski:2016} argue that long-term stability in this system is more likely for planets in the resonant configuration.
Amongst our MR set of 200 simulations of Kepler-60 analogs, we found (by visual examination of the resonant angles) only one case of resonant libration in the resonance chain, and this case was long-term stable.
Given that all of our TTV-mass-based simulations were stable to nearly 100 Myr (and the vast majority of these do not have resonant librations), it is clear that stability on that timescale is not reliant on the resonant configuration, though it is possible that  gigayear stability does depend on resonant libration.
We note that the Exoplanet Archive lists an additional, much longer period, unconfirmed planet candidate in this system; future studies might consider whether a long-period planet could help stabilize the architecture of Kepler-60's inner planets.

\smallskip
\noindent\textit{Kepler-80}: 
Kepler-80 is a five-planet system with multiple librating three-body resonances; the lowest-order librating angles identified by \cite{MacDonald:2016} are $\phi= 3\lambda_b - 5\lambda_e + 2\lambda_d$ and $\phi = 2\lambda_c - 3\lambda_b + \lambda_e$. 
In our simulations of this system architecture, only 2 of the 100 TTV-mass based simulations were stable to 5 billion inner planet orbits ($\sim$13 Myr), and just 10 of the MR relationship simulations were stable. 
From visual inspection of the short integrations for our TTV-based initial conditions, $\sim$25\% showed libration of the 3-body angles above for the resonances between planets c-b-e and planets d-e-b; we note that both stable simulations showed libration.
For the MR relationship simulations, $\sim$30\%  showed libration of those 3-body angles, including 5 of the 10 stable simulations.
None of our simulations showed the potential three-body resonance between planets g-c-b identified by \cite{MacDonald:2021}.
Given the low stability rates in our simulations, and the higher rate of stability in the subset of our simulations that showed libration of the resonances for this system, it seems likely that resonant libration can enhance the stability in this system.

\begin{figure}
    \centering
    \includegraphics{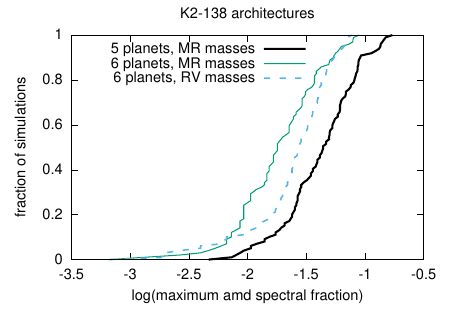}
    \caption{Cumulative distributions of maximum {\sc{amd}} spectral fractions.}
    \label{f:k2-138}
\end{figure}

\smallskip
\noindent\textit{K2-138}: The first five transiting planets in the K2-138 system were reported by \cite{Christiansen:2018} with a sixth planet later reported by \cite{Hardegree-Ullman:2021}. 
Our initial set of simulations for this system was based on MR relationship masses for the initial five planets; this set was almost entirely unstable as only 3 of the 100 simulations survived the 32 Myr simulations. 
We ran a new set of MR simulations including the sixth planet, which has an orbital period of 42 days, significantly longer than the 2-13 day orbital periods of the inner five planets. In this set, the stability rate jumped to just over half of the simulations (54 out of 100 survived to 32 Myr). 
This highlights how the presence of an additional, longer period planet can help stabilize a tightly packed inner planetary system. 
When we examine the {\sc{amd}} spectral fractions of the planets in the 5-planet version of the K2-138 system, they are systematically larger than those in the 6 planet system.
In other words, the presence of the additional longer-period planet tends to cause the secular evolution of all the planets to be more `orderly' with fewer secular frequencies controlling the system's evolution.
Figure~\ref{f:k2-138} shows the maximum {\sc{amd}} spectral fraction for the three versions of the K2-138 system we considered, including the 6-planet version with masses constrained from RV measurements \citep{Lopez:2019}.
The simulations using RV masses had slightly larger spectral fractions than the MR mass simulations despite having a higher survival rate (77 of the 100 RV simulations survive to 32 Myr); however, simulations of the K2-138 system with 6 planets have more cases with significantly smaller spectral fractions than the 5-planet version.
It is possible that other observed architectures that are particularly prone to instability could be similarly stabilized by additional not-yet-observed planets. 
Though we note that any additional outer planets would need to not induce significant inclination evolution that would make transits much less likely \cite[see, e.g., ][]{Becker:2017}.

\smallskip
\noindent\textit{Kepler-23}: For the three observed planets in Kepler-23, we note a significant difference in the stability rate when using TTV and MR derived masses. 
When adopting the TTV masses from \cite{Hadden:2014}, the majority (87\%) of our simulations become unstable within the 97 Myr simulations compared to none of the MR simulations.
The planets in this system are similar sizes, with radii in the range 2-3$R_{\earth}$, so the MR relationship from \cite{Wolfgang:2016} yields masses for all planets of less than $\sim25M_{\earth}$.
The TTV solution from \cite{Hadden:2014} yields mass estimates for planets b and d similar to the MR estimates, but the mass estimate for planet c is $50-70M_{\earth}$.
This large mass for planet c does not seem compatible with long-term stability, suggesting the mass estimate is too large.
This is consistent with the note in Table 2 of \cite{Hadden:2014} that planet c is likely to have an eccentricity large enough to cause the nominal TTV mass to be an overestimate.

\smallskip
\noindent\textit{Kepler-402}: There are four confirmed transiting planets in the Kepler-402 system, with one additional candidate planet. 
In our simulations of the four confirmed planets using MR masses, 50 of the 100 simulations were unstable within 55 Myr. 
This is another system that warrants future investigation to see if the longer-period planet candidate enhances the stability of the observed architecture.

\smallskip
\noindent\textit{Kepler-444}: The five observed sub-Earth radius planets in Kepler-444 are in orbit around one star in a triple star system (see \citealt{Zhang:2023} for the latest stellar system properties). 
Two planets (d and e) are very close to a 5:4 period ratio \citep[e.g.][]{Mills:2017}, but none of the other planets are particularly close to low-order resonant ratios.
Our simulations of this system's architecture using a MR relationship for the planets has a high rate of instability (81 of 100 simulations ended in a collision). 
We find no significant statistical differences of the key architectural parameters (planetary masses, mass-ratios, or initial Hill radius separations) between the stable and unstable instances of our Kepler-444 architectures.
The TTV-derived masses of planets d and e \citep{Mills:2017} are on the small end of our MR mass ranges, but we find stable simulations spanning the full MR range for these planets, which includes masses 2-3 times more massive than the TTV-derived limits.
We note that the \cite{Mills:2017} TTV measurements are not listed in the exoplanet archive and thus were not included in our simulation initial conditions. 
The MR masses for the other planets span the mass ranges suggested by \cite{Stalport:2022} to be consistent with stability based on a frequency analysis of their shorter-term semimajor axis evolution; \cite{Stalport:2022} also include the effects of the Kepler-444BC binary star around Kepler-444A in their analysis.
Like most Kepler multi-planet systems, the Kepler-444 planetary system is estimated to be quite old (11 Gyr; \citealt{Campante:2015}), implying the real planets are stable on timescales much longer than our $\sim50$Myr integrations despite most of those integrations ending in instability.
Possibly unseen planets or the influence of the companion stars play a role in stabilizing the real system.

\smallskip
\noindent\textit{Kepler-1542}: There are four confirmed transiting planets in the Kepler-1542 system. 
We ran simulations of this system using the original reported planetary radii \citep{Morton:2016} for the MR masses. 
We found that 77 of the 100 simulations were unstable within 40 Myr. 
\cite{Berger:2018} reported planetary radii for this system that are consistently $\sim20$\% larger than those reported earlier, so we ran a second set of simulations with MR masses based on the updated radii which resulted in 70 of the 100 simulations being unstable; the stability rate is not meaningfully improved  with the updated radii.
The Exoplanet Archive lists one additional, slightly larger period unconfirmed planet candidate for this system; a future investigation might consider whether such an additional planet would improve system stability.

\smallskip
\noindent\textit{K2-384}: There are five confirmed transiting planets in the K2-384 system \citep{Christiansen:2022}. 
Our simulations assuming MR masses for the planets show an instability rate similar to the 5-planet version of K2-138 above (75 of the 100 simulations end in collisions). 
The presently known planets have orbital periods ranging from 2-14 days.
We highlight this system as one where the dynamical architecture could point to an additional, stabilizing planet that is presently undetected and could be detected in future observations.

\section{Results and Discussion} \label{sec:discussion}

From our analysis above, we can identify a few factors that appear to be important in determining long-term stability: 
(1) In most of the unstable simulations, the majority of neighboring planets have period ratios below 2 (see, e.g., Figure~\ref{fig:pr_pairs}), which increases the chances that planetary mean motion resonances can overlap at moderate to large eccentricities. 
(2) The unstable simulations have distributions of dynamical separations of neighboring planets that skew toward smaller separations (though $\sim30$\% of collisions in these systems occur between planets at initial separations larger than 20 Hill radii).
(3) Most of the unstable simulations have larger spectral fractions, meaning that the system's secular evolution is not dominated by a small number of secular modes; this increases the chances that secular chaos can increase eccentricities over time. 
(4) The unstable simulations tend to have slightly larger coefficients of variation in planetary masses, meaning the planets are less likely to have very similar masses; this trend is weak, but large inter-planet mass variations might contribute to larger secular eccentricity increases.

We can explore the relative influence of these planetary system features for predicting long-term stability using machine learning.
We trained and tested a simple gradient boosting classifier\footnote{We used the GradientBoostingClassifier algorithm from the scikit-learn python package \citep{scikit-learn}} described at \url{https://scikit-learn.org/stable/modules/generated/sklearn.ensemble.GradientBoostingClassifier.html\#sklearn.ensemble.GradientBoostingClassifier}. For simplicity, we left all input parameters set to their default values. 
using the entire ensemble of simulations that survived long enough for spectral fractions to be calculated ($\sim20000$ simulations, of which $\sim2000$ were unstable on less than $5\times10^9$ orbits of the innermost planet). 
We split our simulation set into a training set consisting of 60\% of the simulations and a testing set of the remaining 40\%, and trained the classifier to predict stability based on different combinations of data features from the simulation set; the training and testing sets remained fixed for all combinations of data features.
To characterize the performance of the classifier, we use three parameters: 1) the positive predictive value (PPV) of the classifier predicting a system is unstable (i.e., the fraction of systems predicted to be unstable that are actually unstable); 2) the classifier's true positive rate (TPR) for identifying unstable systems (i.e., the fraction of all unstable systems correctly classified as unstable); and 3) the area under the receiver operating characteristic curve (ROC-AUC), a summary statistic based on a binary classifier's true positive and false positive rates that is 1 for a perfect classifier and 0.5 for random chance.
We first tested the predictive power of several data features individually: the largest {\sc{amd}} spectral fraction amongst the planets in a simulation, the coefficient of variation for the masses of the planets, the average period ratio for neighboring planets, and the average dynamical separation of neighboring planets.
The PPV, TPR, and ROC-AUC for each of these single-feature classifiers is given in Table~\ref{t:ML}.

The {\sc{amd}} spectral fraction and average period ratio both performed much better than average $\Delta$ or mass $C_v$.
Training a classifier using the combination of {\sc{amd}} spectral fraction and average period ratio yields a PPV of 0.79, TPR of 0.67, and ROC-AUC of 0.82; adding mass $C_v$ and average $\Delta$ results in very small improvements.  
When we add in a variety of other easily calculated data features (system-wide minimum $\Delta$, minimum and maximum period ratios, the coefficients of similarity and variation for mass and radius, the maximum spectral fractions for $a$, $e$, and $i$) we get modest improvements in the classifier's performance: a PPV of 0.81, TPR of 0.78, and ROC-AUC of 0.88. 
Including our assessment of whether planets in these systems are in, near, or not near MMRs does not change the classifier's performance in these metrics.
This supports the idea that MMRs are not always directly driving the instabilities.
As noted in Section~\ref{ss:sim-summary}, the near-resonant systems do have a higher instability rate, but do not account for the majority of the instabilities. 
However, MMRs may play a role in destabilizing systems that experience eccentricity growth due to secular interactions, especially in the systems with smaller period ratios.

\begin{deluxetable}
{l|c|c|c}
\tablecaption{The positive predictive value (PPV), true positive rate (TPR) and area under the receiver operating characteristic curve (ROC-AUC) for  machine learning classifiers trained using the listed planetary system parameters to make a binary prediction for whether a system is stable or unstable.\label{t:ML}}
\tablehead{system features   & PPV & TPR & ROC-AUC }
\startdata
{\sc{amd}} spectral fraction & 0.65 & 0.54 & 0.75 \\
average period ratio & 0.83 & 0.47 & 0.73 \\
above 2 features & 0.79 & 0.67 & 0.82\\
average $\Delta$ & 0.66 & 0.06 & 0.53 \\
$C_V$ mass & 0.66 & 0.03 & 0.51 \\ 
above 4 features & 0.80 & 0.69 & 0.84\\
plus additional features\tablenotemark{*}  & 0.81 & 0.78 & 0.88 \\
\enddata
\tablenotetext{*}{system-wide minimum $\Delta$, minimum and maximum period ratios, the coefficients of similarity and variation for mass and radius, the maximum spectral fractions for $a$, $e$, and $i$. }
\end{deluxetable}

Overall, our machine learning analysis shows that the {\sc {amd}} spectral fraction is a particularly useful and easily calculated instability indicator alongside the period ratios and dynamical separations of nearest-neighbor planets.
The importance of the {\sc {amd}} spectral fraction in predicting (in)stability supports the finding in \cite{Volk:2020} that many instabilities seen in simulations can be attributed to secular chaos.
There are additional potential drivers of instability in multi-planet systems that warrant additional investigations, and there remain many insights to be gained from detailed studies of stability as a function of planetary spacings, masses, and orbital period distributions (see, e.g., some recent examples: \citealt{Tamayo:2020,Petit:2020,Tamayo:2021,Lissauer:2021,Goldberg:2022,Rath:2022,Obertas:2023,Rice:2023, Dietrich:2023}).

\section{Conclusions} \label{sec:conclusions}

We have analysed a large set of simulated planetary systems with three or more planets based on architectures of observed exoplanet systems. 
Systems that experience instabilities within $5\times10^9$ orbits of the innermost planet tend to share the following properties:
\begin{itemize}
    \item The secular evolution is not dominated by just a small number of secular frequencies, but instead has many frequencies in the evolution of the planets' eccentricities and inclinations.
    \item The system includes one or more adjacent planet pairs of period ratio below two.
    \item Smaller planetary separations (measured in units of mutual Hill radius) and more unequal planetary masses are more common, though they are not necessary for instability.
\end{itemize}
Spectral fractions that characterize the secular evolution of a system based on very short integrations can provide a prediction of long-term stability, especially when combined with other slightly weaker predictive features such as period ratios, planet dynamical spacings, and characterizations of the mass distribution.
These stability predictors might allow dynamical refinement of mass estimates for some observed planets or identify systems whose stability could be enhanced by as-yet-undetected additional planets.

\facilities{Exoplanet Archive, ADS}
\software{rebound,numpy \citep{2020Natur.585..357H},scikit-learn}

{\it Acknowledgements:}
This work was supported by NASA (grant 80NSSC18K0397); KV acknowledges additional support from NASA grant 80NSSC20K0267.
The results reported herein benefited from collaborations and/or information exchange within the program ``Alien Earths'' (supported by the National Aeronautics and Space Administration under Agreement No. 80NSSC21K0593) for NASA’s Nexus for Exoplanet System Science (NExSS) research coordination network sponsored by NASA’s Science Mission Directorate.
This research has made use of the NASA Exoplanet Archive, which is operated by the California Institute of Technology, under contract with the National Aeronautics and Space Administration under the Exoplanet Exploration Program.
An allocation of computer time from the UA Research Computing High Performance Computing (HPC) at the University of Arizona is gratefully acknowledged.
This paper includes data collected by the Kepler and K2 missions. Funding for the Kepler and K2 missions is provided by the NASA Science Mission directorate.

\appendix
\section{System details}\label{a:systems}

 Table~\ref{tab:systems} lists details for every set of 100 simulations we ran (described in Section~\ref{sec:simulations}). 
Note that many systems appear more than once in the table; for example, systems with TTV/RV planet mass constraints appear once for the simulation set based on those constraints and once for the simulations set based on MR relationship planet masses. 
Some systems also appear with differing numbers of planets. 
These are systems for which we ran an initial simulation set based on data available in 2019 but have subsequently had additional planets detected in the system. 

\startlongtable
\begin{deluxetable*}
{|p{0.095\linewidth}|p{0.03\linewidth}|p{0.0675\linewidth}|p{0.055\linewidth}|p{0.6\linewidth}|}
\tablecaption{List of system architectures used in this work.\label{tab:systems}}
    \tablehead{system name & n$_{pl}$ & num. of unstable sims & MMRs$^a$ & notes }
\startdata
Kepler-90	&	8	&	98	&	2	&	MR planet masses except for h,f; upper limits on masses for planets h, f from  \cite{Santerne:2016};  planet radii except i from \cite{Cabrera:2014}; planet i radius from \cite{Shallue:2018}; stellar mass from \cite{Santerne:2016} \\
Kepler-90	&	8	&	88	&	2	&	MR planet masses except for h,f; planet f mass from alternate MR relationship; upper limit for planet h mass from  \cite{Santerne:2016};  planet radii except i from \cite{Cabrera:2014}; planet i radius from \cite{Shallue:2018}; stellar mass from \cite{Santerne:2016} \\
K2-138	&	6	&	23	&	2	&	RV planet masses, planet radii, and stellar mass from \cite{Lopez:2019}\\
K2-138	&	6	&	46	&	2	&	MR planet masses; planet radii, and stellar mass from \cite{Lopez:2019}\\
K2-138	&	5*	&	97	&	2	&	MR planet masses; planet radii and stellar mass from \cite{Christiansen:2018} \\
Kepler-11	&	6	&	85	&	2	&	TTV planet masses and eccentricities for b-f, planet radii, and stellar mass from \cite{Lissauer:2013}; MR mass for planet g\\
Kepler-11	&	6	&	34	&	2	&	MR planet masses except for planet e; planet radii, TTV mass for planet e, and stellar mass from \cite{Lissauer:2013} \\
Kepler-11	&	6	&	3	&	2	&	TTV planet masses for all planets except g, TTV eccentricities for planets d, e, f, planet radii and stellar mass from \cite{Lissauer:2013}; MR mass for planet g \\
Kepler-20	&	6	&	100	&	2	&	RV masses for planets b,c, d,  g, eccentricities for planets b, c, g, and radii for  planets b, c, d, e, f  from \cite{Buchhave:2016}; MR masses for planets e and f; planet g does not transit and its radius is taken from the Exoplanet Archive's estimate based on a mass-radius; relationship; stellar mass from \cite{Morton:2016}; un-modeled stellar companion \\
Kepler-20	&	6	&	20	&	2	&	MR masses for all planets except g; planet radii and RV mass for planet g from \cite{Buchhave:2016}; planet g does not transit and its radius is taken from the Exoplanet Archive's estimate based on a mass-radius; stellar mass from \cite{Morton:2016}; un-modeled stellar companion \\
Kepler-80	&	6	&	98	&	1	&	TTV planet masses for b, c, d, e from \cite{MacDonald:2016}; MR planet f mass; planet radii and stellar mass from \cite{MacDonald:2016} \\
Kepler-80	&	6	&	90	&	1	&	MR planet masses; planet radii and stellar mass from \cite{MacDonald:2016} \\
K2-268	&	5	&	1	&	2	&	MR planet masses; planet radii and stellar mass from \cite{deLeon:2021}\\
K2-384	&	5	&	75	&	1	&	MR planet masses; planet radii and stellar mass from  \cite{Christiansen:2022}  \\
Kepler-102	&	5	&	92	&	2	&	RV planet masses for d, e from \cite{Marcy:2014}; MR planet masses for b, c, f;  planet radii from \cite{Marcy:2014}; stellar mass from \cite{Wang:2014} \\
Kepler-102	&	5	&	79	&	2	&	MR planet masses; planet radii from \cite{Marcy:2014}; stellar mass from \cite{Wang:2014} \\
Kepler-122	&	5	&	2	&	0	&	MR planet masses (mass for planet c from alternate MR relationship); planet radii from \cite{Rowe:2014}; stellar mass from \cite{Morton:2016} \\
Kepler-150	&	5	&	0	&	0	&	MR planet masses; planet radii from \cite{Rowe:2014}; stellar mass from \cite{Morton:2016} \\
Kepler-154	&	5	&	1	&	0	&	MR planet masses; planet radii for b, c from \cite{Rowe:2014}; planet radii for d, e, f   and stellar mass from \cite{Morton:2016} \\
Kepler-169	&	5	&	9	&	2	&	MR planet masses; planet radii from \cite{Rowe:2014}; stellar mass from \cite{Morton:2016} \\
Kepler-186	&	5	&	63	&	0	&	MR planet masses; planet b, c, d, e radii from \cite{Rowe:2014}; planet f radius and eccentricity from \cite{Torres:2015};  stellar mass from \cite{Morton:2016} \\
Kepler-186	&	5	&	0	&	0	&	MR planet masses; planet b, c, d, e radii from \cite{Rowe:2014}; planet f radius from \cite{Torres:2015};  stellar mass from \cite{Morton:2016} \\
Kepler-238	&	5	&	0	&	0	&	TTV planet masses and radii for e, f from \cite{Xie:2014}; MR planet masses for b, c, d; radii for b, c, d from \cite{Rowe:2014}; stellar mass from \cite{Morton:2016} \\
Kepler-238	&	5	&	0	&	0	&	MR planet masses for all planetx except e; planet e mass from \cite{Xie:2014}; radii for b, c, d from \cite{Rowe:2014}; radii for e,f from \cite{Xie:2014}; stellar mass from \cite{Morton:2016} \\
Kepler-292	&	5	&	9	&	0	&	MR planet masses; planet radii from \cite{Rowe:2014}; stellar mass from \cite{Morton:2016} \\
Kepler-296	&	5	&	1	&	0	&	MR planet masses; planet radii and stellar mass from \cite{Barclay:2015}; unmodelled stellar companion \\
Kepler-32	&	5	&	1	&	0	&	MR planet masses; planet radii and stellar mass from \cite{Fabrycky:2012}\\
Kepler-32	&	5	&	0	&	0	&	TTV planet masses for b, c from \cite{Hadden:2014}; MR planet masses for d, e, f; planet radii and stellar mass from \cite{Fabrycky:2012}\\
Kepler-33	&	5	&	43	&	2	&	TTV masses for all planets except b from \cite{Hadden:2016}; MR mass for planet b; planet radii and stellar mass from \cite{Lissauer:2012} \\
Kepler-444	&	5	&	100	&	1	&	MR planet masses; planet eccentricities, radii, and stellar mass from \cite{Campante:2015}; there are two unmodelled stellar companions \\
Kepler-444	&	5	&	81	&	1	&	MR planet masses; planet radii, and stellar mass from \cite{Campante:2015}; there are two unmodelled stellar companions  \\
Kepler-55	&	5	&	0	&	1	&	MR planet masses; planets d, e, f radii from \cite{Rowe:2014}; planets b, c radii  from \cite{Steffen:2013}; stellar mass from \cite{Morton:2016} \\
Kepler-55	&	5	&	0	&	1	&	TTV planet b, c masses from \cite{Hadden:2014}; MR planet masses for d, e, f; planets d, e, f radii from \cite{Rowe:2014}; planets b, c radii  from \cite{Steffen:2013}; stellar mass from \cite{Morton:2016} \\
Kepler-62	&	5	&	0	&	0	&	MR planet masses; planet radii and stellar mass from  \cite{Borucki:2013}  \\
Kepler-82	&	5	&	1	&	0	&	TTV planet masses and radii for b, c, f from \cite{Freudenthal:2019}; MR planet masses for d, e; planet d, e, radii from \cite{Berger:2018}; l stellar mass from \cite{Freudenthal:2019}\\
Kepler-82	&	4*	&	0	&	0	&	MR planet masses for d, e; TTV planet masses for b, c from \cite{Hadden:2014}; planets b, c radii from \cite{Xie:2013}; planets d, e radii from \cite{Rowe:2014}; stellar mass from \cite{Morton:2016}\\
Kepler-84	&	5	&	0	&	0	&	MR planet masses; planets b, c radii from \cite{Xie:2014}; planets d, e, f radii from \cite{Rowe:2014}; stellar mass from \cite{Morton:2016} \\
Kepler-84	&	5	&	0	&	0	&	TTV planet masses for b, c from \cite{Hadden:2014}; MR planet d, e, f masses; planets b, c radii from \cite{Xie:2014}; planets d, e, f radii from \cite{Rowe:2014}; stellar mass from \cite{Morton:2016} \\
K2-187	&	4	&	0	&	0	&	MR planet masses; planet radii and stellar mass from \cite{Mayo:2018} \\
K2-266	&	4	&	13	&	1	&	TTV planet masses for d, e from \cite{Rodriguez:2018}; MR planet masses for b, c; planet radii and stellar mass from \cite{Rodriguez:2018}; un-modeled stellar companion \\
K2-266	&	4	&	9	&	1	&	MR planet masses; planet radii and stellar mass from \cite{Rodriguez:2018};   un-modeled stellar companion \\
K2-285	&	4	&	0	&	2	&	MR planet masses; planet radii and stellar mass from  \cite{Palle:2019}  \\
K2-285	&	4	&	6	&	2	&	RV planet masses for b, c, planet radii, and stellar mass from  \cite{Palle:2019}; MR masses for d, e  \\
K2-32	&	4	&	2	&	0	&	RV planet masses for all planets and eccentricities for planets d, e, f, planet radii, and stellar mass from \cite{Lillo-Box:2020}; unmodeled stellar companion\\
K2-32	&	4	&	0	&	0	&	MR planet masses; planet radii and stellar mass from \cite{Lillo-Box:2020}; unmodeled stellar companion\\
K2-72	&	4	&	6	&	0	&	MR planet masses; planet radii and stellar mass from \cite{Dressing:2017} \\
Kepler-106	&	4	&	1	&	0	&	MR planet masses; planet radii and stellar mass from \cite{Marcy:2014} \\
Kepler-106	&	4	&	0	&	0	&	RV planet masses, planet radii, and stellar mass from \cite{Marcy:2014}\\
Kepler-107	&	4	&	87	&	2	&	MR planet masses; eccentricities from \cite{VanEylen:2015}; planet radii from \cite{Rowe:2014};  stellar mass from \cite{Morton:2016} \\
Kepler-107	&	4	&	8	&	2	&	MR planet masses; planet radii from \cite{Rowe:2014}; stellar mass from \cite{Morton:2016} \\
Kepler-107	&	4	&	6	&	2	&	RV planet masses for b, c, e from \cite{Bonomo:2019}; MR mass for d;  planet radii from\cite{Rowe:2014}; stellar mass from \cite{Morton:2016} \\
Kepler-1388	&	4	&	0	&	0	&	MR planet masses; planet radii and stellar mass from \cite{Morton:2016} \\
Kepler-1542	&	4	&	77	&	2	&	MR planet masses; planet radii and stellar mass from \cite{Morton:2016} \\
Kepler-1542	&	4	&	70	&	2	&	MR planet masses; planet radii from \cite{Berger:2018}; stellar mass from \cite{Morton:2016} \\
Kepler-167	&	4	&	0	&	0	&	MR planet masses (mass for planet e from alternate MR relationship); planet radii and stellar mass from \cite{Kipping:2016};   un-modeled stellar companion \\
Kepler-172	&	4	&	0	&	0	&	MR planet masses; planet radii from \cite{Rowe:2014}; stellar mass from \cite{Morton:2016} \\
Kepler-176	&	4	&	0	&	2	&	MR planet masses; planet b, c, d radii from \cite{Rowe:2014}; planet e radius and stellar mass from \cite{Morton:2016} \\
Kepler-197	&	4	&	100	&	0	&	TTV planet mass for c from \cite{Hadden:2014}; MR masses for planets  b, d, e; eccentricity for planet e from \cite{VanEylen:2015}; planet radii from \cite{Rowe:2014}; setllar mass from \cite{Morton:2016}; un-modeled stellar companion \\
Kepler-197	&	4	&	0	&	0	&	MR planet masses; planet radii from \cite{Rowe:2014}; stellar mass from \cite{Morton:2016}; un-modeled stellar companion \\
Kepler-208	&	4	&	8	&	0	&	MR planet masses; planet radii from \cite{Rowe:2014}; stellar mass from \cite{Morton:2016} \\
Kepler-215	&	4	&	0	&	0	&	MR planet masses; planet radii from \cite{Rowe:2014}; stellar mass from \cite{Morton:2016} \\
Kepler-220	&	4	&	0	&	0	&	MR planet masses; planet radii from \cite{Rowe:2014}; stellar mass from \cite{Morton:2016} \\
Kepler-221	&	4	&	0	&	0	&	MR planet masses; planet radii from \cite{Rowe:2014}; stellar mass from \cite{Morton:2016} \\
Kepler-223	&	4	&	80	&	1	&	TTV planet masses and eccentricities, planet radii, and stellar mass from \cite{Mills:2016} \\
Kepler-223	&	4	&	21	&	1	&	MR planet masses for b, c; TTV planet masses for d, e, planet radii, and stellar mass from \cite{Mills:2016} \\
Kepler-224	&	4	&	0	&	0	&	MR planet masses; planet radii from \cite{Rowe:2014}; stellar mass from \cite{Morton:2016} \\
Kepler-235	&	4	&	0	&	0	&	MR planet masses; planet radii from \cite{Rowe:2014};  stellar mass from \cite{Torres:2017} \\
Kepler-24	&	4	&	6	&	2	&	MR planet masses; radii for planets b, c from \cite{Ford:2012}; radii for planets d, e from \cite{Rowe:2014}; stellar mass from \cite{Morton:2016} \\
Kepler-24	&	4	&	18	&	2	&	TTV planet masses for b, c from \cite{Hadden:2014}; MR planet masses for d, e; planet b, d, c radii from \cite{Berger:2018}; planet e radius and stellar mass from \cite{Morton:2016} \\
Kepler-245	&	4	&	0	&	2	&	MR planet masses; planet b, c, d radii from \cite{Rowe:2014}; planet e radius and stellar mass from \cite{Morton:2016} \\
Kepler-251	&	4	&	0	&	0	&	MR planet masses; planet radii from \cite{Rowe:2014}; stellar mass from \cite{Morton:2016} \\
Kepler-256	&	4	&	0	&	0	&	MR planet masses; planet radii from \cite{Rowe:2014}; stellar mass from \cite{Morton:2016} \\
Kepler-26	&	4	&	0	&	2	&	TTV planet masses for b, c from \cite{Jontof-Hutter:2016}; MR planet masses for d, e; radii for d, e, from \cite{Rowe:2014}; radii for b, c and stellar mass from \cite{Jontof-Hutter:2016} \\
Kepler-26	&	4	&	6	&	2	&	MR planet masses; radii for d, e, from \cite{Rowe:2014}; radii for b, c and stellar mass from \cite{Jontof-Hutter:2016} \\
Kepler-265	&	4	&	0	&	0	&	MR planet masses; radii from \cite{Rowe:2014}; stellar mass from \cite{Batalha:2013} \\
Kepler-282	&	4	&	18	&	0	&	MR planet masses; planets b, c radii from \cite{Rowe:2014}; planets d, e radii from \cite{Xie:2014}; stellar mass from \cite{Morton:2016} \\
Kepler-282	&	4	&	36	&	0	&	TTV planet masses and radii for d, e from \cite{Xie:2014}; MR planet masses for b, c; radii for b, c from \cite{Rowe:2014}; stellar mass from \cite{Morton:2016} \\
Kepler-286	&	4	&	0	&	0	&	MR planet masses; planet radii from \cite{Rowe:2014}; stellar mass from \cite{Morton:2016} \\
Kepler-299	&	4	&	0	&	0	&	MR planet masses; planet radii from \cite{Rowe:2014}; stellar mass from \cite{Morton:2016} \\
Kepler-304	&	4	&	0	&	0	&	MR planet masses; planet b, c, d radii from \cite{Rowe:2014}; planet e radius and stellar mass from \cite{Morton:2016} \\
Kepler-306	&	4	&	0	&	0	&	MR planet masses; planet radii from \cite{Rowe:2014}; stellar mass from \cite{Morton:2016} \\
Kepler-324	&	4	&	0	&	2	&	MR planet masses; planets b, c, radii from \cite{Berger:2018}; planet e radius from \cite{Valizadegan:2022}; planet d radius and stellar mass from \cite{Jontof-Hutter:2021} \\
Kepler-338	&	4	&	79	&	0	&	MR planet masses; eccentricities from \cite{VanEylen:2015}; planet radii from \cite{Rowe:2014};  stellar mass from \cite{Morton:2016} \\
Kepler-338	&	4	&	2	&	0	&	MR planet masses; planet radii from \cite{Rowe:2014};  stellar mass from \cite{Morton:2016} \\
Kepler-341	&	4	&	0	&	0	&	MR planet masses; planet radii from \cite{Rowe:2014}; stellar mass from \cite{Morton:2016} \\
Kepler-342	&	4	&	2	&	1	&	MR planet masses; planets b, c, d radii from \cite{Rowe:2014}; planet e radius and stellar mass from \cite{Morton:2016} \\
Kepler-37	&	4	&	19	&	2	&	MR planet masses; radius for planet b from \cite{Stassun:2017}; planet e radius from Q1-Q8 KOI table; radii for planets d, e and stellar mass from \cite{Marcy:2014}\\
Kepler-402	&	4	&	50	&	2	&	MR planet masses; planet radii from \cite{Rowe:2014}; stellar mass from \cite{Morton:2016} \\
Kepler-411	&	4	&	40	&	2	&	MR planet masses for b, d, e; RV planet c mass, planet radii, and stellar mass from \cite{Sun:2019} \\ 
Kepler-48	&	4	&	0	&	0	&	RV planet masses, radii for planets b, c, d, and stellar mass from \cite{Marcy:2014}; planet e does not transit, so its radius is taken from the Exoplanet Archive's estimate based on a mass-radius relationship  \\
Kepler-48	&	4	&	0	&	0	&	MR planet masses for b, c, d; RV mass for planet e, radii for planets b, c, d,  and stellar mass from \cite{Marcy:2014}; planet e does not transit, so its radius is taken from the Exoplanet Archive's estimate based on a mass-radius relationship  \\
Kepler-49	&	4	&	1	&	1	&	MR planet masses; planets b, c radii from \cite{Steffen:2013}; planets e, d radii from  \cite{Rowe:2014}; stellar mass from \cite{Morton:2016} \\
Kepler-65	&	4	&	0	&		&	RV planet masses and stellar mass from \cite{Mills:2019}; planet radii from \cite{Chaplin:2013}; planet e does not transit, so the radius is taken from the Exoplanet Archive estimate \\
Kepler-65	&	3*	&	0	&	0	&	MR planet masses; planet radii and stellar mass from \cite{Chaplin:2013} \\
Kepler-758	&	4	&	4	&	2	&	MR planet masses; planet radii and stellar mass from  \cite{Morton:2016}  \\
Kepler-79	&	4	&	0	&	0	&	TTV planet masses and eccentricities, planet radii, and stellar mass from \cite{Jontof-Hutter:2014} \\
Kepler-79	&	4	&	0	&	0	&	MR planet masses; planet radii, and stellar mass from \cite{Jontof-Hutter:2014} \\
Kepler-85	&	4	&	95	&	1	&	MR planet masses; planets b, c radii from \cite{Xie:2013}; planets d, e radii from \cite{Rowe:2014}; stellar mass from \cite{Morton:2016} \\
Kepler-85	&	4	&	91	&	1	&	TTV planet c mass from \cite{Hadden:2014}; TTV planet e mass from \cite{Hadden:2017}; MR planet d, d masses; planets b, c radii from \cite{Xie:2013}; planets d, e radii from \cite{Rowe:2014}; stellar mass from \cite{Morton:2016} \\
Kepler-94	&	4	&	89	&	0	&	RV planet masses and eccentricities, and planet radii from \cite{Weiss:2013}; stellar mass from \cite{Albrecht:2013}; unmodeled stellar companion\\
Kepler-94	&	4	&	0	&	0	&	RV planet masses  for planets c, d, e, and planet radii from \cite{Weiss:2013}; MR mass for planet b; stellar mass from \cite{Albrecht:2013}; unmodeled stellar companion\\
K2-133	&	3	&	0	&	0	&	MR planet masses; planet radii and stellar mass from \cite{Wells:2018} \\
K2-136	&	3	&	0	&	0	&	MR planet masses; planet radii and stellar mass from \cite{Mann:2018};   un-modeled stellar companion \\
K2-148	&	3	&	8	&	2	&	MR planet masses; planet radii and stellar mass from \cite{Hirano:2018}; un-modeled stellar companion \\
K2-155	&	3	&	0	&	0	&	MR planet masses; planet radii and stellar mass from \cite{DiezAlonso:2018} \\
K2-165	&	3	&	0	&	0	&	MR planet masses; planet radii and stellar mass from \cite{Mayo:2018} \\
K2-183	&	3	&	0	&	0	&	MR planet masses; planet radii and stellar mass from \cite{Mayo:2018} \\
K2-19	&	3	&	0	&	1	&	RV masses for planets b,c and stellar masses from \cite{Petigura:2020}; MR mass for planet d; planet radii from \cite{Sinukoff:2016} \\
K2-219	&	3	&	0	&	2	&	MR planet masses; planet radii and stellar mass from \cite{Mayo:2018} \\
K2-233	&	3	&	0	&	0	&	MR planet masses; planet radii and stellar mass from \cite{David:2018} \\
K2-3	&	3	&	0	&	0	&	RV planet masses from \cite{Damasso:2018}; planet radii from \cite{Crossfield:2016}; stellar mass from \cite{Kosiarek:2019} \\
K2-3	&	3	&	0	&	0	&	MR planet masses; planet radii from \cite{Damasso:2018}; stellar mass from \cite{Kosiarek:2019} \\
K2-37	&	3	&	1	&	0	&	MR planet masses; planet radii from \cite{Sinukoff:2016}; stellar mass from \cite{Crossfield:2016} \\
K2-58	&	3	&	0	&	0	&	MR planet masses; planet radii and stellar mass from \cite{Crossfield:2016} \\
K2-80	&	3	&	0	&	2	&	MR planet masses; planet radii for c, b and stellar mass from \cite{Crossfield:2016};  planet d radius from \cite{Mayo:2018}  \\
Kepler-100	&	3	&	0	&	0	&	MR planet masses for c, d; planet radii, RV mass for planet b, and stellar mass  from \cite{Marcy:2014} \\
Kepler-104	&	3	&	0	&	0	&	MR planet masses; planet radii from \cite{Rowe:2014}; stellar mass from  \cite{Morton:2016}; un-modeled stellar companion \\
Kepler-114	&	3	&	0	&	0	&	TTV planet masses for c, d from \cite{Xie:2014}; MR planet mass for b; planet radii from \cite{Rowe:2014}; stellar mass from \cite{Morton:2016} \\
Kepler-114	&	3	&	0	&	0	&	MR planet masses; planet b radius form \cite{Rowe:2014}, planets c and d radii from \cite{Xie:2014}; stellar mass from \cite{Morton:2016} \\
Kepler-124	&	3	&	0	&	0	&	MR planet masses; planet radii from \cite{Rowe:2014}; stellar mass from \cite{Morton:2016} \\
Kepler-1254	&	3	&	0	&	0	&	MR planet masses; planet radii and stellar mass from \cite{Morton:2016} \\
Kepler-126	&	3	&	0	&	0	&	MR planet masses; planet radii from \cite{Rowe:2014}; stellar mass   from \cite{SilvaAguirre:2015} \\
Kepler-127	&	3	&	0	&	0	&	MR planet masses; planet radii from \cite{Rowe:2014}; stellar mass from \cite{Morton:2016} \\
Kepler-130	&	3	&	0	&	0	&	MR planet masses; planet radii from \cite{Rowe:2014};  stellar mass from \cite{Morton:2016}; un-modeled stellar companion \\
Kepler-138	&	3	&	40	&	1	&	TTV planet masses, planet eccentricities, and stellar mass from  \cite{Jontof-Hutter:2015}  \\
Kepler-138	&	3	&	8	&	1	&	MR planet masses; planet radii and stellar mass from  \cite{Jontof-Hutter:2015}  \\
Kepler-142	&	3	&	0	&	0	&	MR planet masses; planet radii from \cite{Rowe:2014}; stellar mass from \cite{Morton:2016} \\
Kepler-157	&	3	&	0	&	0	&	MR planet masses; planet radii for b, c from \cite{Rowe:2014}; planet radius for d and  stellar mass from \cite{Morton:2016} \\
Kepler-164	&	3	&	0	&	0	&	MR planet masses; planet radii from \cite{Rowe:2014}; stellar mass from \cite{Morton:2016} \\
Kepler-166	&	3	&	0	&	0	&	MR planet masses; planet radii for b,c from \cite{Rowe:2014};   planet radius for d and stellar mass from \cite{Morton:2016} \\
Kepler-171	&	3	&	0	&	0	&	MR planet masses; planet radii from \cite{Rowe:2014}; stellar mass from \cite{Morton:2016} \\
Kepler-174	&	3	&	0	&	0	&	MR planet masses; planet radii from \cite{Rowe:2014}; stellar mass from \cite{Morton:2016} \\
Kepler-178	&	3	&	0	&	0	&	MR planet masses; planet radii from \cite{Rowe:2014}; stellar mass from \cite{Morton:2016} \\
Kepler-184	&	3	&	3	&	0	&	MR planet masses; planet radii from \cite{Rowe:2014}; stellar mass from \cite{Morton:2016} \\
Kepler-19	&	3	&	9	&	0	&	RV/TTV planet masses and planet c and d eccentricities from \cite{Malavolta:2017}; planet b radius and stellar mass from \cite{Ballard:2011}; planets c and d do not transit, so radii are taken from the Exoplanet Archive's estimate based on a mass-radius relationship  \\
Kepler-191	&	3	&	0	&	2	&	MR planet masses; planet b, c radii from \cite{Rowe:2014}; planet d radius and stellar mass from \cite{Morton:2016} \\
Kepler-192	&	3	&	0	&	0	&	MR planet masses; planet b, c radii from \cite{Rowe:2014}; planet d radius and stellar mass from \cite{Morton:2016} \\
Kepler-194	&	3	&	0	&	0	&	MR planet masses; planet radii from \cite{Rowe:2014}; stellar mass from \cite{Morton:2016} \\
Kepler-198	&	3	&	0	&	0	&	MR planet masses; planet b, c radii from \cite{Rowe:2014}; planet d radius and stellar mass from \cite{Morton:2016} \\
Kepler-203	&	3	&	0	&	0	&	MR planet masses; planet radii from \cite{Rowe:2014}; stellar mass from \cite{Morton:2016} \\
Kepler-207	&	3	&	0	&	0	&	MR planet masses; planet radii from \cite{Rowe:2014}; stellar mass from \cite{Morton:2016} \\
Kepler-217	&	3	&	3	&	0	&	MR planet masses; planet b, c radii from \cite{Rowe:2014}; planet d radius and stellar mass from \cite{Morton:2016} \\
Kepler-218	&	3	&	0	&	0	&	MR planet masses; planet b, c radii from \cite{Rowe:2014}; planet d radius and stellar mass from \cite{Morton:2016} \\
Kepler-219	&	3	&	0	&	0	&	MR planet masses; planet radii from \cite{Rowe:2014}; stellar mass from \cite{Morton:2016} \\
Kepler-226	&	3	&	32	&	0	&	MR planet masses; planet radii from \cite{Rowe:2014}; stellar mass from \cite{Morton:2016} \\
Kepler-23	&	3	&	87	&	2	&	TTV planet masses from \cite{Hadden:2014}; planet b, c, radii from \cite{Ford:2012}; planet d radius from \cite{Rowe:2014}; stellar mass from \cite{Morton:2016} \\
Kepler-23	&	3	&	0	&	2	&	MR planet masses; planet b, c, radii from \cite{Ford:2012}; planet d radius from \cite{Rowe:2014}; stellar mass from \cite{Morton:2016} \\
Kepler-244	&	3	&	0	&	0	&	MR planet masses; planet radii from \cite{Rowe:2014}; stellar mass from \cite{Morton:2016} \\
Kepler-249	&	3	&	0	&	0	&	MR planet masses; planet radii from \cite{Rowe:2014}; stellar mass from \cite{Morton:2016} \\
Kepler-25	&	3	&	0	&	0	&	RV planet masses, radii for planets b, c and stellar mass from \cite{Marcy:2014}; planet d does not transit, so its radius is taken from the Exoplanet Archive's estimate based on a mass-radius relationship; unmodeled stellar companion  \\
Kepler-25	&	3	&	0	&	0	&	MR planet masses for b, c; radii for planets b, c, RV mass for planet d, and stellar mass from \cite{Marcy:2014}; planet d does not transit, so its radius is taken from the Exoplanet Archive's estimate based on a mass-radius relationship; unmodeled stellar companion  \\
Kepler-250	&	3	&	0	&	0	&	MR planet masses; planet radii from \cite{Rowe:2014}; stellar mass from \cite{Morton:2016} \\
Kepler-253	&	3	&	0	&	0	&	MR planet masses; planet radii from \cite{Rowe:2014}; stellar mass from \cite{Morton:2016} \\
Kepler-254	&	3	&	1	&	1	&	MR planet masses; planet radii from \cite{Rowe:2014}; stellar mass from \cite{Morton:2016} \\
Kepler-255	&	3	&	0	&	0	&	MR planet masses; planet b, c radii from \cite{Rowe:2014}; planet d radius and stellar mass from \cite{Morton:2016} \\
Kepler-267	&	3	&	0	&	0	&	MR planet masses; planet radii from \cite{Rowe:2014}; stellar mass from \cite{Morton:2016} \\
Kepler-271	&	3	&	3	&	0	&	MR planet masses; planet radii, planet orbital periods, and stellar mass from \cite{Morton:2016}; note that conflicting solutions for planet b's orbital period have been published, and there are two additional unconfirmned and\ unmodeled planet candidates in this system \\
Kepler-272	&	3	&	0	&	0	&	MR planet masses; planet radii from \cite{Rowe:2014}; stellar mass from \cite{Morton:2016} \\
Kepler-275	&	3	&	0	&	0	&	MR planet masses; planet radii from \cite{Rowe:2014}; stellar mass from \cite{Morton:2016} \\
Kepler-276	&	3	&	0	&	0	&	MR planet masses; planet b radius from \cite{Rowe:2014}; planets c, d radii from \cite{Xie:2014}; stellar mass from \cite{Morton:2016} \\
Kepler-288	&	3	&	0	&	0	&	MR planet masses; planet radii from \cite{Rowe:2014}; stellar mass from \cite{Morton:2016} \\
Kepler-289	&	3	&	0	&	0	&	TTV planet masses and planet radii from \cite{Schmitt:2014}; stellar mass from \cite{Morton:2016} \\
Kepler-289	&	3	&	0	&	0	&	MR masses for planets b, d; TTV mass for planet c and planet radii from \cite{Schmitt:2014}; stellar mass from \cite{Morton:2016} \\
Kepler-295	&	3	&	0	&	0	&	MR planet masses; planet radii from \cite{Rowe:2014}; stellar mass from \cite{Morton:2016} \\
Kepler-298	&	3	&	0	&	0	&	MR planet masses; planet radii from \cite{Rowe:2014}; stellar mass from \cite{Morton:2016} \\
Kepler-30	&	3	&	0	&	0	&	TTV planet masses and eccentricities and planet radii from \cite{Sanchis-Ojeda:2012}; stellar mass from \cite{Fabrycky:2012} \\
Kepler-30	&	3	&	0	&	0	&	MR mass for planet b; TTV planet masses for c, d and planet radii from \cite{Sanchis-Ojeda:2012}; stellar mass from \cite{Fabrycky:2012} \\
Kepler-301	&	3	&	0	&	0	&	MR planet masses; planet radii from \cite{Rowe:2014}; stellar mass from \cite{Morton:2016} \\
Kepler-305	&	3	&	0	&	1	&	MR planet masses; planets b, c radii from \cite{Xie:2014}; planet d radius from \cite{Rowe:2014}; stellar mass from \cite{Morton:2016}\\
Kepler-310	&	3	&	0	&	0	&	MR planet masses; planet radii from \cite{Rowe:2014}; stellar mass from \cite{Morton:2016} \\
Kepler-319	&	3	&	0	&	0	&	MR planet masses; planet radii from \cite{Rowe:2014}; stellar mass from \cite{Morton:2016} \\
Kepler-325	&	3	&	0	&	0	&	MR planet masses; planet radii from \cite{Rowe:2014}; stellar mass from \cite{Morton:2016} \\
Kepler-326	&	3	&	1	&	2	&	MR planet masses; planet radii from \cite{Rowe:2014}; stellar mass from \cite{Morton:2016} \\
Kepler-327	&	3	&	0	&	0	&	MR planet masses; planet radii from \cite{Rowe:2014}; stellar mass from \cite{Morton:2016} \\
Kepler-331	&	3	&	0	&	0	&	MR planet masses; planet radii from \cite{Rowe:2014}; stellar mass from \cite{Morton:2016} \\
Kepler-332	&	3	&	0	&	0	&	MR planet masses; planet radii from \cite{Rowe:2014}; stellar mass from \cite{Morton:2016} \\
Kepler-334	&	3	&	0	&	2	&	MR planet masses; planet radii from \cite{Rowe:2014}; stellar mass from \cite{Morton:2016} \\
Kepler-336	&	3	&	0	&	0	&	MR planet masses; planet radii from \cite{Rowe:2014}; stellar mass from \cite{Morton:2016} \\
Kepler-339	&	3	&	20	&	2	&	MR planet masses; planet radii from \cite{Rowe:2014}; stellar mass from \cite{Morton:2016} \\
Kepler-350	&	3	&	4	&	0	&	TTV planet masses and planet radii for c, d from \cite{Xie:2014}; MR planet mass for b; planet b radius from \cite{Rowe:2014}; stellar mass from \cite{Morton:2016} \\
Kepler-350	&	3	&	0	&	0	&	MR planet masses; planet radii for c, d from \cite{Xie:2014}; planet b radius from \cite{Rowe:2014}; stellar mass from \cite{Morton:2016} \\
Kepler-351	&	3	&	2	&	0	&	MR planet masses; planet b, c radii from \cite{Rowe:2014}; planet d radius and stellar mass from \cite{Morton:2016} \\
Kepler-354	&	3	&	2	&	0	&	MR planet masses; planet radii from \cite{Rowe:2014}; stellar mass from \cite{Morton:2016} \\
Kepler-357	&	3	&	0	&	0	&	MR planet masses; planet radii from \cite{Rowe:2014}; stellar mass from \cite{Morton:2016} \\
Kepler-363	&	3	&	0	&	0	&	MR planet masses; planet radii from \cite{Rowe:2014}; stellar mass from \cite{Morton:2016} \\
Kepler-372	&	3	&	0	&	1	&	MR planet masses; planet radii from \cite{Rowe:2014}; stellar mass from \cite{Morton:2016} \\
Kepler-374	&	3	&	0	&	0	&	MR planet masses; planet radii from \cite{Rowe:2014}; stellar mass from \cite{Morton:2016} \\
Kepler-398	&	3	&	0	&	0	&	MR planet masses; planet b, c radii from \cite{Rowe:2014}; planet d radius and stellar mass from \cite{Morton:2016} \\
Kepler-399	&	3	&	0	&	0	&	MR planet masses; planet radii from \cite{Rowe:2014}; stellar mass from \cite{Morton:2016} \\
Kepler-401	&	3	&	0	&	0	&	MR planet masses; planet b, c radii from \cite{Rowe:2014}; planet d radius and stellar mass from \cite{Morton:2016} \\
Kepler-403	&	3	&	0	&	0	&	MR planet masses; planet b, c radii from \cite{Rowe:2014}; planet d radius and stellar mass from \cite{Morton:2016} \\
Kepler-42	&	3	&	0	&	0	&	MR planet masses; planet radii and stellar mass from \cite{Muirhead:2012} \\
Kepler-431	&	3	&	7	&	0	&	MR planet masses; planet radii and stellar mass from \cite{Everett:2015} \\
Kepler-445	&	3	&	32	&	1	&	MR planet masses; planet radii and stellar mass from \cite{Muirhead:2015} \\
Kepler-446	&	3	&	0	&	0	&	MR planet masses; planet radii and stellar mass from \cite{Muirhead:2015} \\
Kepler-450	&	3	&	0	&	0	&	MR planet masses (mass for planet b from alternate MR relationship); planet radii from \cite{VanEylen:2015}; stellar mass from \cite{Huber:2013} \\
Kepler-52	&	3	&	0	&	0	&	MR planet masses; planet b, c radii from \cite{Rowe:2014}; planet d radius and stellar mass from \cite{Morton:2016} \\
Kepler-52	&	3	&	0	&	0	&	TTV planet b, c masses from \cite{Hadden:2014}; MR mass for planet d; planet radii and stellar mass from \cite{Morton:2016} \\
Kepler-53	&	3	&	0	&	0	&	MR planet masses; planet d radius from \cite{Rowe:2014}; planets b, c radii and stellar mass from \cite{Steffen:2013} \\
Kepler-53	&	3	&	0	&	0	&	TTV planet b, c masses from \cite{Hadden:2014}; MR mass for planet d; planet d radius from \cite{Rowe:2014}; planets b, c radii and stellar mass from \cite{Steffen:2013} \\
Kepler-54	&	3	&	5	&	1	&	MR planet masses; planet d radius from \cite{Rowe:2014}; planets b, c radii  from \cite{Steffen:2013}; stellar mass from \cite{Morton:2016} \\
Kepler-54	&	3	&	21	&	1	&	TTV planet b, c masses from \cite{Hadden:2014}; MR mass for planet d; planet d radius from \cite{Rowe:2014}; planets b, c radii and stellar mass from \cite{Morton:2016} \\
Kepler-58	&	3	&	0	&	0	&	MR planet masses; planet d radius from \cite{Rowe:2014}; planets b, c radii  from \cite{Steffen:2013}; stellar mass from \cite{Morton:2016} \\
Kepler-60	&	3	&	61	&	1	&	MR planet masses; planet radii and stellar mass from  \cite{Jontof-Hutter:2016}  \\
Kepler-60	&	3	&	0	&	1	&	TTV planet masses, planet radii, and stellar mass from  \cite{Jontof-Hutter:2016}  \\
Kepler-770	&	3	&	0	&	0	&	MR planet masses; planet radii and stellar mass from  \cite{Morton:2016}  \\
Kepler-81	&	3	&	0	&	2	&	MR planet masses; planets b, c radii from \cite{Rowe:2014}; planets d, e radii from \cite{Xie:2014}; stellar mass from \cite{Morton:2016} \\
Kepler-81	&	3	&	0	&	2	&	TTV planets b,c masses from \cite{Hadden:2014}; MR mass for planet d; planets b, c radii from \cite{Rowe:2014}; planets d, e radii from \cite{Xie:2014}; stellar mass from \cite{Morton:2016} \\
Kepler-83	&	3	&	0	&	0	&	MR planet masses; planets b, c radii from \cite{Xie:2013}; planet d radius from \cite{Rowe:2014}; stellar mass from \cite{Morton:2016} \\
Kepler-83	&	3	&	0	&	0	&	TTV planet masses from \cite{Hadden:2014}; planets b, c radii from \cite{Xie:2013}; planet d radius from \cite{Rowe:2014}; stellar mass from \cite{Morton:2016} \\
Kepler-9	&	3	&	0	&	1	&	TTV masses and radii for planets b, c, from \cite{Holman:2010}; MR planet d mass; planet d radius and stellar mass from \cite{Torres:2011} \\
Kepler-92	&	3	&	86	&	0	&	TTV planet masses and eccentricities for b, c from \cite{Xie:2014}; MR mass for planet d; planets b, c radii from \cite{Xie:2014}; planet d radius from \cite{VanEylen:2015}; stellar mass from \cite{SilvaAguirre:2015} \\
Kepler-92	&	3	&	0	&	0	&	MR planet masses; planets b, c radii from \cite{Xie:2014}; planet d radius from \cite{VanEylen:2015}; stellar mass from \cite{SilvaAguirre:2015} \\
\enddata
    \tablecomments{If the value of $n_{pl}$ for a system is marked with an asterisk, at least one additional planet was detected in that system since its data was queried for the initial set of simulations; in these cases the system has a second entry in the table/simulation set with the additional planet included. In all cases except one (Kepler-271; see notes for that system), the orbital periods for all planets were taken to be the default value and uncertainties in the Exoplanet Archive composite planet table\footnote{\url{https://exoplanetarchive.ipac.caltech.edu/cgi-bin/TblView/nph-tblView?app=ExoTbls&config=PSCompPars}}; in almost all cases, the orbital periods for each planet from different lightcurve analyses agree within uncertainties. Unless otherwise indicated in the notes column, inclinations and eccentricities for all planets were assigned from the Rayleigh distributions described in Section~\ref{sec:simulations}. MR masses are generated from the statistical mass-radius relationship from \cite{Wolfgang:2016} for planets with radii of 4$R_{\earth}$ or smaller. For planets exceeding that size limit, we either use a TTV/RV mass measurement (indicated in the notes column) or we use the mass and uncertainty given in the Exoplanet Archive Composite data table, which is based on the mass-radius relationship from \cite{Chen:2017}, see \url{https://exoplanetarchive.ipac.caltech.edu/docs/composite_calc.html}; for systems where we used this mass estimate, the specific planet is indicated in the notes column as coming from an alternate MR relationship.\\
    $^a$0 indicates no planets are near strong MMRs (mean motion resonances); 1 indicates one or more pair of planets is in or very close to a strong MMR; 2 indicates one or more pair of planets is close enough to a strong MMR that resonant interactions cannot entirely be ruled out (especially if simulated planets evolve to large eccentricities).\\}
\end{deluxetable*}

\end{document}